\documentclass[twocolumn,times,tighten]{aastex62}
\usepackage{newtxmath}
\usepackage{natbib}

\usepackage{epsfig}
\usepackage{amsmath,amssymb}
\usepackage{ulem}
\usepackage{graphicx}

\usepackage{comment}

\begin{document}

\renewcommand*{\sectionautorefname}{\S\!}
\renewcommand*{\subsectionautorefname}{\S\!}
\renewcommand*{\subsubsectionautorefname}{\S\!}

\font\sevenrm=cmr7 scaled 1000

\newcommand{\Msun}{\ifmmode M_\odot \else $M_\odot$\fi}
\newcommand{\vabs}{$15,500$ km s$^{-1}$}
\newcommand{\swift}{\textsl{Swift}}
\newcommand{\hst}{\textsl{HST}}
\newcommand{\gptf}{$g_\textrm{PTF}$}
\newcommand{\rptf}{$R_\textrm{PTF}$}
\newcommand{\Mbh}{\ifmmode M_\text{BH} \else $M_\text{BH}$\fi}
\newcommand{\RNum}[1]{\uppercase\expandafter{\romannumeral #1\relax}}

\newcommand{\cmsq}{\ifmmode {\text{cm}^{-2}} \else cm$^{-2}$\fi}
\newcommand{\kms}{\ifmmode {\text{km~s}^{-1}} \else km~s$^{-1}$\fi}

\newcommand{\La}{\ifmmode {\rm Ly}\alpha \else Ly$\alpha$\fi}
\newcommand{\Ka}{\ifmmode {\rm K}\alpha \else K$\alpha$\fi}
\newcommand{\Lb}{\ifmmode {\rm L}\beta \else L$\beta$\fi}
\newcommand{\Ha}{\ifmmode {\rm H}\alpha \else H$\alpha$\fi}
\newcommand{\Hb}{\ifmmode {\rm H}\beta \else H$\beta$\fi}
\newcommand{\Hgamma}{\ifmmode {\rm H}\gamma \else H$\gamma$\fi}
\newcommand{\Hdelta}{\ifmmode {\rm H}\delta \else H$\delta$\fi}
\newcommand{\Pa}{\ifmmode {\rm P}\alpha \else P$\alpha$\fi}
\newcommand{\CIIIb}{\ifmmode {\rm C}\,{\sc iii]}\,\lambda1909
    \else C\,{\sc iii]}\,$\lambda1909$\fi}
\newcommand{\CIV}{\ifmmode {\rm C}\,{\sc iv}\,\lambda1549
    \else C\,{\sc iv}\,$\lambda1549$\fi}

\newcommand{\OVI}{\ifmmode {\rm O}\,{\sc vi}\,\lambda1035
    \else O\,{\sc vi}\,$\lambda1035$\fi}
\newcommand{\HI}{\ifmmode {\rm H}\,{\sc i}
    \else H\,{\sc i}\fi}
\newcommand{\HeIs}{\ifmmode {\rm He}\,{\sc i}\,^*\lambda3889
    \else He\,{\sc i}\,$^*\lambda3889$\fi}
\newcommand{\HeIss}{\ifmmode {\rm He}\,{\sc i}\,^*\lambda10830
    \else He\,{\sc i}\,$^*\lambda10830$\fi}
\newcommand{\HeI}{\ifmmode {\rm He}\,{\sc i}\,^*\lambda5876
    \else He\,{\sc i}\,$^*\lambda5876$\fi}
\newcommand{\HeII}{\ifmmode {\rm He}\,{\sc ii}\,\lambda4686
    \else He\,{\sc ii}\,$\lambda4686$\fi}
\newcommand{\HeIIa}{\ifmmode {\rm He}\,{\sc ii}\,\lambda1640
    \else He\,{\sc ii}\,$\lambda1640$\fi}
\newcommand{\HeIIb}{\ifmmode {\rm He}\,{\sc ii}\,\lambda1085
    \else He\,{\sc ii}\,$\lambda1085$\fi}
\newcommand{\CII}{\ifmmode {\rm C}\,{\sc ii}\,\lambda1335
    \else C\,{\sc ii}\,$\lambda1335$\fi}
\newcommand{\CIII}{\ifmmode {\rm C}\,{\sc iii}\,\lambda977
    \else C\,{\sc iii}\,$\lambda977$\fi}

\newcommand{\bOIIIb}{\ifmmode {\rm [O}\,{\sc iii]}\,\lambda5007
    \else [O\,{\sc iii]}\,$\lambda5007$\fi}
\newcommand{\OIIIb}{\ifmmode {\rm O}\,{\sc iii]}\,\lambda1663
    \else O\,{\sc iii]}\,$\lambda1663$\fi}
\newcommand{\OVIII}{\ifmmode {\rm O}\,{\sc viii}\,653~{\rm eV}
    \else O\,{\sc viii}\,$653~{\rm eV}$\fi}
\newcommand{\OVII}{\ifmmode {\rm O}\,{\sc vii}\,568~eV
    \else O\,{\sc vii}\,$568~{\rm eV}$\fi}

\newcommand{\OIVb}{\ifmmode {\rm O}\,{\sc iv]}\,\lambda1402
    \else O\,{\sc iv]}\,$\lambda1402$\fi}
\newcommand{\bOIIb}{\ifmmode {\rm [O}\,{\sc ii]}\,\lambda3727
    \else [O\,{\sc ii]}\,$\lambda3727$\fi}
\newcommand{\bOIb}{\ifmmode {\rm [O}\,{\sc i]}\,\lambda6300
    \else [O\,{\sc i]}\,$\lambda6300$\fi}
\newcommand{\OI}{\ifmmode {\rm [O}\,{\sc i]}\,\lambda1304
    \else [O\,{\sc i]}\,$\lambda1304$\fi}
\newcommand{\NII}{\ifmmode {\rm N}\,{\sc ii}\,\lambda1084
    \else N\,{\sc ii}\,$\lambda1084$\fi}
\newcommand{\NIII}{\ifmmode {\rm N}\,{\sc iii}\,\lambda990
    \else N\,{\sc iii}\,$\lambda990$\fi}
\newcommand{\NIIIb}{\ifmmode {\rm N}\,{\sc iii]}\,\lambda1750
    \else N\,{\sc iii]}\,$\lambda1750$\fi}
\newcommand{\NIVb}{\ifmmode {\rm N}\,{\sc iv]}\,\lambda1486
    \else N\,{\sc iv]}\,$\lambda1486$\fi}
\newcommand{\NV}{\ifmmode {\rm N}\,{\sc v}\,\lambda1240
    \else N\,{\sc v}\,$\lambda1240$\fi}
\newcommand{\MgII}{\ifmmode {\rm Mg}\,{\sc ii}\,\lambda\lambda2796, 2803
      \else Mg\,{\sc ii}\,$\lambda\lambda2796, 2803$\fi}
\newcommand{\CVI}{\ifmmode {\rm C}\,{\sc vi}\,368~eV
      \else C\,{\sc vi}\,368~eV\fi}
\newcommand{\SiIV}{\ifmmode {\rm Si}\,{\sc iv}\,\lambda1397
    \else Si\,{\sc iv}\,$\lambda1397$\fi}
\newcommand{\bFeXb}{\ifmmode {\rm [Fe}\,{\sc x]}\,\lambda6734
      \else [Fe\,{\sc x]}\,$\lambda6734$\fi}
\newcommand{\MgX}{\ifmmode {\rm Mg}\,{\sc x}\,\lambda615
      \else Mg\,{\sc x}\,$\lambda615$\fi}
\newcommand{\MgXI}{\ifmmode {\rm Mg}\,{\sc xi}\,1.34~keV
      \else Mg\,{\sc xi}\,1.34~keV\fi}
\newcommand{\MgXII}{\ifmmode {\rm Mg}\,{\sc xii}\,1.47~keV
    \else Mg\,{\sc xii}\,1.47~keV\fi}
\newcommand{\bNeVb}{\ifmmode {\rm [Ne}\,{\sc v]}\,\lambda3426
    \else [Ne\,{\sc v]}\,$\lambda3426$\fi}
\newcommand{\NeVIII}{\ifmmode {\rm Ne}\,{\sc viii}\,\lambda774
    \else Ne\,{\sc viii}\,$\lambda774$\fi}
\newcommand{\SiVIIa}{\ifmmode {\rm Si}\,{\sc vii}\,\lambda70
    \else Si\,{\sc vii}\,$\lambda70$\fi}
\newcommand{\NeVIIa}{\ifmmode {\rm Ne}\,{\sc vii}\,\lambda88
    \else Ne\,{\sc vii}\,$\lambda88$\fi}
\newcommand{\NeVIIIa}{\ifmmode {\rm Ne}\,{\sc viii}\,\lambda88
    \else Ne\,{\sc viii}\,$\lambda88$\fi}
\newcommand{\NeIX}{\ifmmode {\rm Ne}\,{\sc ix}\,915~eV
    \else Ne\,{\sc ix}\,915~eV\fi}
\newcommand{\NeX}{\ifmmode {\rm Ne}\,{\sc x}\,1.02~keV
    \else Ne\,{\sc x}\,1.02~keV\fi}
\newcommand{\SiII}{\ifmmode {\rm Si}\,{\sc ii}\,\lambda1265
    \else Si\,{\sc ii}\,$\lambda1265$\fi}
\newcommand{\SiIII}{\ifmmode {\rm Si}\,{\sc iii}\,\lambda1206
    \else Si\,{\sc iii}\,$\lambda1206$\fi}
\newcommand{\SiXII}{\ifmmode {\rm Si}\,{\sc xii}\,\lambda506
      \else Si\,{\sc xii}\,$\lambda506$\fi}
\newcommand{\SiXIII}{\ifmmode {\rm Si}\,{\sc xiii}\,1.85~keV
    \else Si\,{\sc xiii}\,1.85~keV\fi}
\newcommand{\SiXIV}{\ifmmode {\rm Si}\,{\sc xiv}\,2.0~keV
    \else Si\,{\sc xiv}\,2.0~keV\fi}
\newcommand{\SXV}{\ifmmode {\rm S}\,{\sc xv}\,2.45~keV
    \else S\,{\sc xv}\,2.45~keV\fi}
\newcommand{\SXVI}{\ifmmode {\rm S}\,{\sc xvi}\,2.62~keV
    \else S\,{\sc xvi}\,2.62~keV\fi}
\newcommand{\ArXVII}{\ifmmode {\rm Ar}\,{\sc xvii}\,3.10~keV
    \else Ar\,{\sc xvii}\,3.10~keV\fi}
\newcommand{\ArXVIII}{\ifmmode {\rm Ar}\,{\sc xviii}\,3.30~keV
    \else Ar\,{\sc xviii}\,3.30~keV\fi}
\newcommand{\FeIII}{\ifmmode {\rm Fe}\,{\sc iii}
    \else Fe\,{\sc iii}\fi}
\newcommand{\FeIXVI}{\ifmmode {\rm Fe}\,{\sc 1-16}\,6.4~keV
    \else Fe\,{\sc 1-16}\,6.4~keV\fi}
\newcommand{\FeXVIIXXIII}{\ifmmode {\rm Fe}\,{\sc 17-23}\,6.5~keV
    \else Fe\,{\sc 17-23}\,6.5~keV\fi}
\newcommand{\FeXXV}{\ifmmode {\rm Fe}\,{\sc xxv}\,6.7~keV
    \else Fe\,{\sc xxv}\,6.7~keV\fi}
\newcommand{\FeXXVI}{\ifmmode {\rm Fe}\,{\sc xxvi}\,6.96~keV
    \else Fe\,{\sc xxvi}\,6.96~keV\fi}
\newcommand{\FeLa}{\ifmmode {\rm Fe}\,{\sc L}\,0.7-0.8~keV
    \else Fe\,{\sc L}\,0.7-0.8~keV\fi}
\newcommand{\FeLb}{\ifmmode {\rm Fe}\,{\sc L}\,1.03-1.15~keV
    \else Fe\,{\sc L}\,1.03-1.15~keV\fi}
\newcommand{\AlIII}{\ifmmode {\rm Al}\,{\sc iii}\,\lambda1857
    \else Al\,{\sc iii}\,$\lambda1857$\fi}

\title{\uppercase{Discovery of Highly Blueshifted Broad Balmer and Metastable Helium Absorption Lines in a \\
Tidal Disruption Event}}   
\author[0000-0002-9878-7889]{T.\ Hung}
\affiliation{Department of Astronomy and Astrophysics, University of
California, Santa Cruz, CA 95064, USA}

\author[0000-0003-1673-970X]{S.~B. Cenko}
\affiliation{Astrophysics Science Division, NASA Goddard Space Flight Center, MC 661, Greenbelt, MD 20771, USA}
\affiliation{Joint Space-Science Institute, University of Maryland, College Park, MD 20742, USA}

\author[0000-0002-6485-2259]{Nathaniel Roth}
\affiliation{Joint Space-Science Institute, University of Maryland, College Park, MD 20742, USA}
\affiliation{Department of Astronomy, University of Maryland, College Park, MD  20742, USA}

\author[0000-0003-3703-5154]{S. Gezari}
\affiliation{Department of Astronomy, University of Maryland, College Park, MD  20742, USA}
\affiliation{Joint Space-Science Institute, University of Maryland, College Park, MD 20742, USA}

\author{S. Veilleux}
\affiliation{Institute of Astronomy and Kavli Institute for Cosmology Cambridge}
\affiliation{University of Cambridge, Cambridge CB3 0HA, United Kingdom}

\author[0000-0002-3859-8074]{Sjoert van Velzen}
\affiliation{Department of Astronomy, University of Maryland, College Park, MD 20742}
\affiliation{Center for Cosmology and Particle Physics, New York University, NY 10003}

\author{C.~Martin Gaskell}
\affiliation{Department of Astronomy and Astrophysics, University of
California, Santa Cruz, CA 95064, USA}

\author{Ryan J. Foley}
\affiliation{Department of Astronomy and Astrophysics, University of
California, Santa Cruz, CA 95064, USA}

\author[0000-0003-0901-1606]{N.\ Blagorodnova}
\affiliation{Department of Astrophysics/IMAPP, Radboud University, Nijmegen, The Netherlands}

\author[0000-0003-1710-9339]{Lin Yan}
\affiliation{Caltech Optical Observatories, California Institute of Technology, Pasadena, CA 91125, USA}

\author[0000-0002-3168-0139]{M.~J. Graham}
\affiliation{Division of Physics, Mathematics, and Astronomy, California Institute of Technology, Pasadena, CA 91125, USA}

\author{J.~S. Brown}
\affiliation{Department of Astronomy and Astrophysics, University of
California, Santa Cruz, CA 95064, USA}

\author{M.~R. Siebert}
\affiliation{Department of Astronomy and Astrophysics, University of
California, Santa Cruz, CA 95064, USA}

\author[0000-0001-9676-730X]{Sara Frederick}
\affiliation{Department of Astronomy, University of Maryland, College Park, MD 20742, USA}

\author{Charlotte Ward}
\affiliation{Department of Astronomy, University of Maryland, College Park, MD 20742, USA}

\author{Pradip Gatkine}
\affiliation{Department of Astronomy, University of Maryland, College Park, MD 20742, USA}

\author{Avishay Gal-Yam}
\affiliation{Department of Particle Physics and Astrophysics, Weizmann Institute of Science, 234 Herzl St, Rehovot 76100, Israel}

\author{Yi Yang}
\affiliation{Department of Particle Physics and Astrophysics, Weizmann Institute of Science, 234 Herzl St, Rehovot 76100, Israel}

\author{S. Schulze}
\affiliation{Department of Particle Physics and Astrophysics, Weizmann Institute of Science, 234 Herzl St, Rehovot 76100, Israel}

\author{G. Dimitriadis}
\affiliation{Department of Astronomy and Astrophysics, University of
California, Santa Cruz, CA 95064, USA}

\author{Thomas Kupfer}
\affiliation{Kavli Institute for Theoretical Physics, University of California, Santa Barbara, CA 93106, USA}

\author{David L. Shupe}
\affiliation{IPAC, California Institute of Technology, 1200 E. California
             Blvd, Pasadena, CA 91125, USA}

\author{Ben Rusholme}
\affiliation{IPAC, California Institute of Technology, 1200 E. California
             Blvd, Pasadena, CA 91125, USA}

\author{Frank J. Masci}
\affiliation{IPAC, California Institute of Technology, 1200 E. California
             Blvd, Pasadena, CA 91125, USA}

\author{Reed Riddle}
\affiliation{Caltech Optical Observatories, California Institute of Technology, Pasadena, CA 91125, USA}

\author[0000-0001-6753-1488]{Maayane T. Soumagnac}
\affiliation{Benoziyo Center for Astrophysics, Weizmann Institute of Science, Rehovot, Israel}

\author{J. van Roestel}
\affiliation{California Institute of Technology, 1200 East California Boulevard, CA, USA}

\author{Richard Dekany}
\affiliation{Caltech Optical Observatories, California Institute of Technology, Pasadena, CA 91125, USA}

\begin{abstract}

We report the discovery of non-stellar hydrogen Balmer and metastable helium
absorption lines accompanying a transient, high-velocity (0.05$c$) broad
absorption line (BAL) system in the optical spectra of the tidal disruption
event (TDE) AT2018zr ($z=0.071$). In the \hst\ UV spectra, absorption of high-
and low-ionization lines are also present at this velocity, making AT2018zr
resemble a low-ionization broad absorption line (LoBAL) QSO. We conclude that
these transient absorption features are more likely to arise in fast outflows
produced by the TDE than absorbed by the unbound debris. In accordance with
the outflow picture, we are able to reproduce the flat-topped \Ha\ emission in
a spherically expanding medium, without invoking the typical prescription of
an elliptical disk. We also report the appearance of narrow ($\sim$1000~\kms)
N\,{\sc iii}\,$\lambda$4640~\AA, \HeII, \Ha, and \Hb\, emission in the
late-time optical spectra of AT2018zr, which may be a result of UV continuum
hardening at late time as observed by
\swift. Including AT2018zr, we find a high association rate (3 out of 4) of
BALs in the UV spectra of TDEs. This suggests that outflows may be ubiquitous
among TDEs and may be less sensitive to viewing angle effects compared to QSO
outflows.


\end{abstract}

\section{Introduction}

Occasionally, a star passing too close to a black hole may be disrupted by
tidal stresses. This results in an observable transient flare of radiation
powered by the accretion of about half of the stellar debris onto the black
hole \citep{Rees1988,Phinney1989}.
A significant amount of theoretical work has been devoted to predicting the
rate at which the stellar debris fall back to the pericenter. For black holes
with $\Mbh \lesssim 10^7~\Msun$, the fallback rate, and the accretion rate if
circularization is efficient, may exceed the Eddington rate at early times
\citep{Evans1989, 1999ApJ...514..180U, Strubbe2009, 2011MNRAS.410..359L,2016MNRAS.461..948M, 2018MNRAS.478.3016W}, leading to the formation of radiation powered winds or
jets.

Observations across all wavelengths have revealed that outflows may be
ubiquitous among TDEs. In the well-studied TDE ASASSN-14li, highly ionized
outflows have been detected at both low \citep[a few $\times
100~\kms$;][]{2015Natur.526..542M} and high velocities
\citep[$\sim$0.2$c$;][]{2018MNRAS.474.3593K} in the X-ray. Radio observations
of ASASSN-14li also revealed the presence of a sub-relativistic outflow
\citep{Alexander2016} or an off-axis relativistic jet
\citep{vanVelzen2016,2018ApJ...856....1P}.

Spectroscopy is a powerful tool for probing the kinematics and physical
conditions in TDEs. Especially in the far-UV (FUV), spectroscopy can shed
light on the ionization structures owing to the wealth of atomic transitions
encompassed in this wavelength range. For the three TDEs that were observed
with the Space Telescope Imaging Spectrograph (STIS) onboard the
\textit{Hubble Space Telescope} (\hst), blueshifted absorption lines at FUV
wavelengths that signify the presence of outflows were detected in all three
sources, namely, ASASSN-14li, iPTF16fnl, and iPTF15af \citep{Cenko2016,
2018MNRAS.473.1130B, 2018arXiv180907446B}. These absorption lines are thought
to be ``intrinsic'', meaning that the absorbing gas is physically close to the
TDE. In particular, the broad, saturated absorption troughs of high-ionization
transitions in iPTF15af are reminiscent of those seen in broad absorption line
(BAL) QSOs. \cite{2018arXiv180907446B} concluded that these features could
only form in absorbers with high column densities $N_H > 10^{23}~\cmsq$.

Although the current sample of UV spectroscopy of TDEs is small, a few
spectroscopic distinctions between TDEs and QSOs have emerged in the past few
years. For example, common quasar emission lines such as \MgII\ and \CIIIb\
are either weak or entirely missing in TDE spectra. The absence of \MgII\ may be
explained if these TDEs have a hotter continuum that has photoionized most Mg
to higher ionization states \citep{Cenko2016, 2018MNRAS.473.1130B}. On the
other hand, the prominent \NIIIb\ relative to \CIIIb\ in TDEs may imply
abundance anomalies due to CNO cycle in the pre-disrupted star
\citep{Cenko2016, 2016MNRAS.458..127K}. Simulations have shown that the
anomalous abundance features should be present after the time of peak fallback
rate ($t_\text{peak}$) and are more significant in higher mass stars
\citep{2016MNRAS.458..127K, 2017ApJ...846..150Y, 2018ApJ...857..109G}.

Observationally, TDEs have been discovered at a rate of 1--2 per year.
Starting in 2018, we expect to see an order of magnitude increase in TDE
discovery rate from the combined yield of ground-based optical surveys
\citep[e.g.][]{Hung2018}. In light of this opportunity, we have started a
monitoring campaign to obtain a series of UV spectra of newly discovered TDEs
with \hst. AT2018zr (aka PS18kh) is the first target of this campaign (Program
ID 15331; PI Cenko).
In this paper, we present the analysis of intensive spectroscopic observations
of the TDE AT2018zr spanning across the UV and optical wavelengths in the
first three months since discovery, plus two late time ($\Delta t = 169$ d and
248 d) optical spectra.

While inspecting the data, we discovered a high-velocity ($\sim$0.05$c$) broad
absorption line (BAL) system that is accompanied by non-stellar hydrogen
Balmer and metastable helium absorptions, the first time this has been
observed in a TDE. The high S/N optical spectra enabled us to propose that a
spherically expanding outflow is preferred over the elliptical disk model
\citep{2018arXiv180802890H} for generating the observed flat-topped \Ha\ line.

This paper is structured as follows. We summarize the discovery and
photometric observations of AT2018zr in \autoref{sec:at2018zr_discovery},
detail the observation configurations and data reduction for the UV and
optical spectra in this work in \autoref{sec:observations}, describe our
analysis of the emission and absorption features identified in the data in
\autoref{sec:analysis}, and compare these observations with other TDEs and
discuss the results in \autoref{sec:discussion}. We summarize our findings in
\autoref{sec:conclusion}.

\section{Discovery of AT2018zr}
\label{sec:at2018zr_discovery}

AT2018zr (aka PS18kh) is a tidal disruption event first discovered by
Pan-STARRS1 on UT 2018 Mar 02 \citep{2018ATel11473....1T,2018arXiv180802890H}. The flare is
coincident (offset $\lesssim$ 0.1\arcsec) with the galaxy SDSS
J075654.53+341543.6 at a redshift of $z=0.071$
(\autoref{subsec:optical_observations}). Archival observations of this galaxy
suggest the galaxy is dominated by an old stellar population ($t_\text{age} =
9.8$~Gyr) with a stellar mass of $5\times 10^9~\Msun$
\citep{2019apJ...872..198V}. The lack of X-ray emission prior to the TDE flare
suggests little or no AGN activity \citep{2018arXiv180802890H}. A black hole
mass of $\approx$\,$10^7~\Msun$ is inferred from the host photometry
\citep{2018arXiv180802890H, 2019apJ...872..198V}. While the Zwicky Transient Facility \citep[ZTF;][]{2019PASP..131a8002B,2019arXiv190201945G} was still in the
commissioning phase, the survey serendipitously observed this object since
2018 Feb 7. A complete set of ZTF light curves of this source can be found in
\cite{2019apJ...872..198V}.

\cite{2018arXiv180802890H} and \cite{2019apJ...872..198V} analyzed the
broadband multi-wavelength (UV/optical/X-ray/radio) properties of AT2018zr.
Before the object was Sun-constrained, they found the UV and optical emission
of this source corresponds to a constant blackbody temperature of $T \approx
1.4 \times 10^4$~K in the first 40 days then increased to $T \approx 2.2
\times 10^4$~K \citep{2018arXiv180802890H, 2019apJ...872..198V}. A weak, thermal
($kT$\,$\sim$\,100~eV) X-ray source two orders of magnitude less luminous than the
UV was detected \citep{2019apJ...872..198V}. The late-time ($\Delta t \gtrsim$ 170 days) UV/optical
photometric observations show a significant increase in the blackbody
temperature ($T \gtrsim 5 \times 10^4$~K) while the X-ray flux remained almost
the same \citep{2019apJ...872..198V}.

\cite{2018arXiv180802890H} also analyzed the optical spectra of AT2018zr. Their observations were made in the period
before the target became sun-constrained, while ours extend to later epochs
when the target was visible again. Their analysis used a combination of wind,
elliptical disk, and spiral arm to fit the \Ha\ line profile, and inferred a
large size for the accretion disk ($r_\text{in} \sim 500\,r_g$ and
$r_\text{out} \sim 15,\!000\,r_g$).

\section{Observations}
\label{sec:observations}

After it was confirmed that AT2018zr was bright in the UV from \swift\
\citep[Neil Gehrels Swift Observatory;][]{gcg+04} observations, we triggered a
series of spectroscopic follow-up observations with \hst\ STIS as well as other
ground-based optical telescopes. No observations were made when the target
went behind the Sun between 2018 June and 2018 August. Nevertheless, we
resumed following up this source when it became visible again in 2018
September. All the spectra presented in this paper have been corrected for
galactic extinction using the \cite{2011ApJ...737..103S} dust map. Assuming a
\cite{1989ApJ...345..245C} extinction curve, using $R_\text{C}$ = 3.1 and
$E(B-V) = 0.0404 \pm 0.0006$ at this position corresponds to a galactic visual
extinction of $A_\text{V} = 0.124$~mag. Throughout this paper, we adopt a flat
$\Lambda$CDM cosmology with $H_0 = 69.3~\kms$~Mpc$^{-1}$, $\Omega_m = 0.29$,
and $\Omega_\Lambda = 0.71$. The time difference ($\Delta t$) is expressed in
rest-frame time with respect to the $r$ band maximum at MJD 58194.49.

\subsection{HST STIS Spectra}

We obtained 5 epochs of UV spectra of AT2018zr with \hst\ STIS (GO 15331; PI
Cenko) on 2018 April 11 ($\Delta t$=23 d), 25 ($\Delta t$=36 d), 30 ($\Delta t$=41 d), May 20 ($\Delta t$=59 d), and 23 ($\Delta t$=62 d). The spectra were obtained
through a 52\arcsec\ $\times$ 0.\!\arcsec 2 aperture. For the near-UV (NUV)
and FUV MAMA detectors, the G140L and G230L gratings were used in order to
cover the spectral range of 1570--3180~\AA\ and 1150--1730~\AA\ at a
resolution of 1.2~\AA\ and 3.2~\AA, respectively. During each visit, the
observation was obtained over 3 \hst\ orbits, with 3 equal exposures of 674~s in
the NUV and 6 equal exposures of 920~s in the FUV. We combined the
1-dimensional spectra for each epoch using inverse-variance weighting.

\subsection{Optical Spectra}
\label{subsec:optical_observations}

We obtained 14 optical spectra of AT2018zr in total. The observing
configuration for each spectrum is detailed in \autoref{tab:obs_spec}. The
data obtained with the Spectral Energy Distribution Machine (SEDm), an
integral-field-unit (IFU) spectrograph, were automatically processed by the
data reduction pipeline and were flux-calibrated with the observations of
spectrophotometric standard stars \citep{2018PASP..130c5003B}.

Spectra obtained with the Auxiliary-port CAMera (ACAM) on the 4.2-m William
Herschel Telescope (WHT), the Deveny spectrograph on the Discovery Channel
Telescope (DCT), the Double Beam Spectrograph (DBSP) on the Palomar 200-inch
(P200) telescope, and with Gemini/GMOS-N were reduced with standard IRAF
routines. We performed bias subtraction and flat-fielding in the raw science
frames and extracted the 1D spectrum. Afterwards, we performed wavelength and
flux calibration using observations of the arc lamp and flux standard stars
from the same night.

Data obtained with the Keck Low-Resolution Imaging Spectrometer (LRIS) \citep{1995PASP..107..375O} were reduced with the LRIS automated reduction pipeline \footnote{\url{http://www.astro.caltech.edu/~dperley/programs/lpipe.
html}}.

We measured a redshift of $z=0.071\pm0.001$ using Ca~H+K\,$\lambda\lambda$3969, 3934
and the near infrared Ca II triplet (8498, 8542 and 8662 \AA) absorption lines that originated from the host galaxy from the late-time Keck spectrum obtained in December 2018.

\begin{deluxetable*}{lccccc}[!ht]
\tabletypesize{\footnotesize}
\tablewidth{0pt}
\tablecolumns{6}
\tablecaption{Observing details of the optical spectra of AT2018zr\label{tab:obs_spec}}
\tablehead{\colhead{Obs Date} &  \colhead{Phase (days)} & \colhead{Telescope + Instrument} &   \colhead{Slit Width} &  \colhead{Grism/Grating} & \colhead{Exp Time (s)}}
\startdata
      2018-03-07 &    -10 &             P60 + SEDm &               &   N/A     &   1800   \\
      2018-03-26 &      8 &             P60 + SEDm &               &   N/A     &   1800   \\
      2018-03-27 &      9 &             P60 + SEDm &               &   N/A     &   2700   \\
      2018-03-28 &     10 &             WHT + ACAM &  1.0\arcsec\  &   V400    &   3200   \\
      2018-04-04 &     16 &           DCT + DeVeny &  1.5\arcsec\  &   300g/mm &   2400   \\
      2018-04-11 &     23 &           DCT + DeVeny &  1.5\arcsec\  &   300g/mm &   2400   \\
      2018-04-17 &     28 &           Keck1 + LRIS &  1.0\arcsec\  &  400/3400 + 400/8500  &    1250  \\
      2018-04-19 &     30 &        Gemini + GMOS-N &   1.0\arcsec\ &   B600  &   600    \\
      2018-05-05 &     45 &        Gemini + GMOS-N &   1.0\arcsec\ &   B600    &  600    \\
      2018-05-06 &     46 &           DCT + DeVeny &   1.5\arcsec\ &   300g/mm &   3000   \\
      2018-05-10 &     50 &           Keck1 + LRIS &   1.0\arcsec\ &  600/4000 + 400/8500 &          \\
      2018-05-13 &     53 &           DCT + DeVeny &   1.5\arcsec\ &   300g/mm &   1800   \\
      2018-05-19 &     58 &           DCT + DeVeny &   1.5\arcsec\ &   300g/mm &   900   \\
      2018-09-14 &    169 &           Keck1 + LRIS &   1.0\arcsec\ &  600/4000 + 400/8500&     1200 \\
      2018-12-08 &    248 &           Keck1 + LRIS &   1.0\arcsec\ &  600/4000 + 400/8500&     3600 \\
\enddata
\tablenotetext{2}{The Keck Low-Resolution Imaging Spectrometer \citep{1995PASP..107..375O}.}
\end{deluxetable*}

\section{Analysis}
\label{sec:analysis}

The interpretation of the UV and optical spectra are complicated by the
presence of a high velocity transient BAL system at
$v \approx 1.55 \times 10^4~\kms$ (\autoref{subsec:absorptions}).
This BAL system manifests in UV and optical absorption troughs on top of the
TDE continuum or, sometimes, the emission lines. The rest-frame optical broad
absorption lines such as the hydrogen Balmer series and the metastable helium
lines are particularly rare even in BALQSOs. In the following subsections, we
carefully account for these absorption features while identifying and
measuring the UV and optical emission lines.

\subsection{Absorption Lines}
\label{subsec:absorptions}

\begin{figure*}[p]
\centering
\includegraphics[height=7.75in, angle=0]{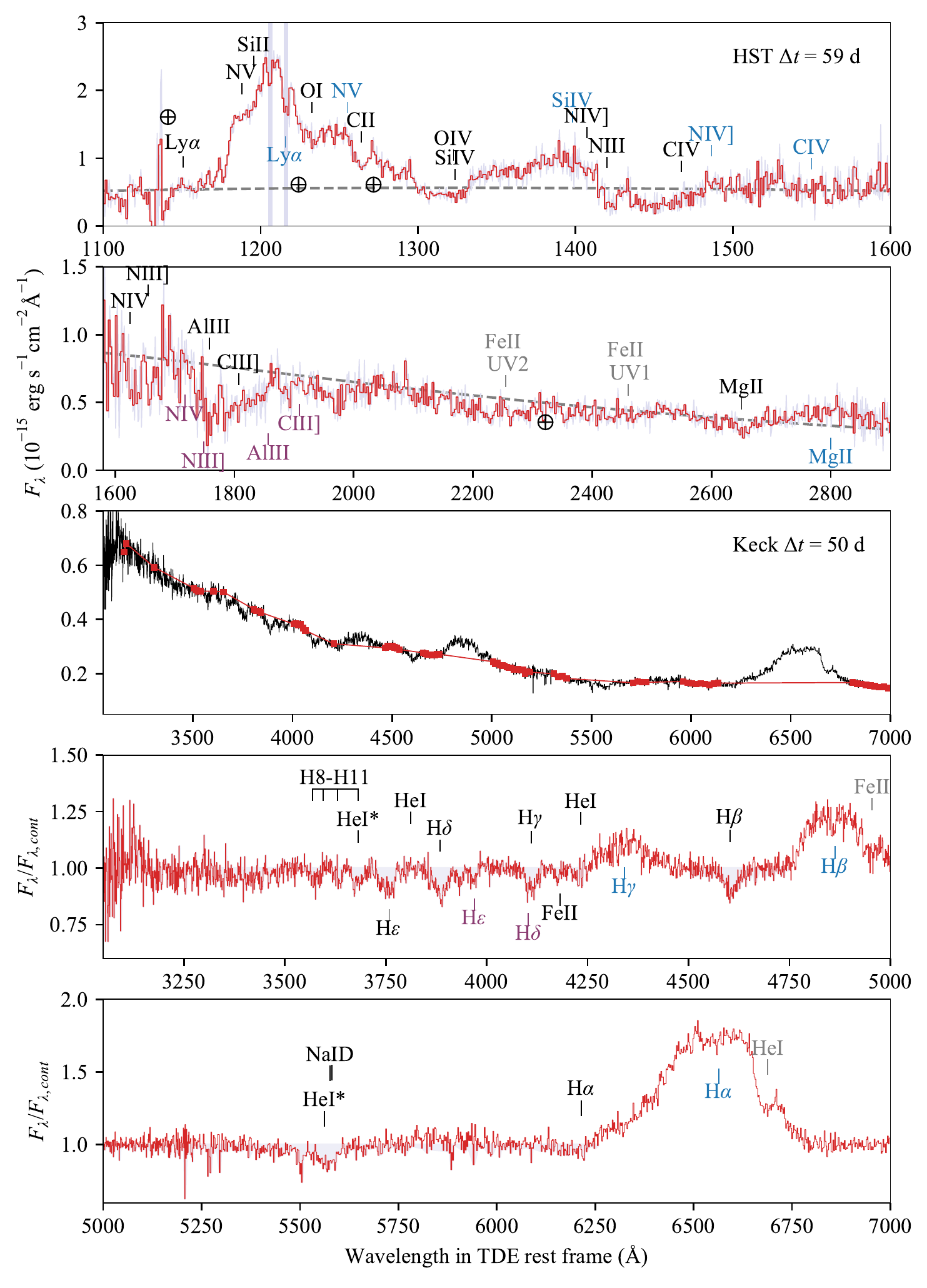}
\caption{Zoom-in view of the UV and optical spectra with high-velocity
($v=15,\!500$~\kms) absorption features indicated by black (detected) and grey
(not detected) labels. The top two panels show the \hst\ spectrum observed on
$\Delta t$=59 d. Since it is hard to cleanly define the continuum in the UV,
we draw a $T=22000$~K blackbody spectrum in grey dashed line to guide the eye.
The wavelengths of the geocoronal airglow lines are indicated by $\oplus$.
Contamination from the host galaxy (\SiIII\ and \La) are shown by the vertical
grey bands. We also mark emission lines at the rest wavelength of the TDE in
blue (detected) and purple (not detected). The middle panel shows the
host-subtracted Keck spectrum from $\Delta t$=50 d, which has the widest
wavelength coverage in the optical. We define the continuum by manually
selecting the line-free regions, as marked by the red squares, and
interpolating linearly between these regions. The normalized spectrum is shown
in the two bottom panels.}
\label{fig:normalized}
\end{figure*}

We identified highly blueshifted hydrogen Balmer series, \HeIs, and \HeI\
transitions in the Keck optical spectrum from $\Delta t$=50 d. These features can be easily seen when the spectrum is normalized with respect to the continuum, as shown in \autoref{fig:normalized}.
We modelled these lines with a single Gaussian and present the EW and FWHM measurements for the hydrogen
and helium absorption lines in \autoref{tab:eqw}. Because the H Balmer and He\,{\sc i} absorption lines are often
accompanied by contamination from nearby features, the
definition of local continuum is more uncertain. For example, the EW measured for
the \Ha\ absorption should be viewed as a lower limit since the blueshifted
\Ha\ absorption is close to the broad blue wing of the \Ha\ emission. Given the blueshifted \Hgamma\ absorption may be susceptible to \Hdelta\ emissions
in the TDE rest frame, the measured EW should also be considered as a lower limit.
We measured a FWHM of 2720$\pm$200~\kms\ for \Hb\ absorption, which is the strongest blueshifted optical absorption line that is free from contamination. From these lines, we derived a mean velocity of (15,500$\pm$400)~\kms\ by employing the relativistic Doppler equation.
Absorptions at this velocity also match with the troughs
seen in the \hst\ STIS spectra in both high (\NV, \SiIV, \NIVb, \CIV, \NIIIb, \CIIIb,
\AlIII) and low (\SiII, \CII, \OI, and \MgII) ionization lines in the last two epochs (day 59 and day 62). We stacked the two \hst\ spectra as there is little spectroscopic evolution during this time, and tabulated a list of detected UV absorption lines in \autoref{tab:uv_absorption}. We also include measurements of the absorption
central wavelength and the width in TDE rest frame, where possible.

\begin{deluxetable}{lccccc}
\tablecaption{Optical absorption features.\label{tab:eqw}}
\tablehead{\colhead{Line} &  \colhead{$\lambda_0$} & \colhead{f} & \colhead{$v_\text{rel}$}& \colhead{EW} & \colhead{FWHM}  \\
 &  (\AA) &  & (\kms) & (\AA) & ($10^3$~\kms) }
\startdata
\Ha &  6564.64 &  0.641080 & $-$16,560 $\pm$ 560 & 2.77$\pm$1.53$^a$ & 2.52 $\pm$ 1.32 \\
HeI &  5875.00 &  0.610230 & $-$15,230 $\pm$ 120 & 9.10$\pm$1.07  & 3.12 $\pm$ 0.21\\
\Hb &  4862.70 &  0.119380 & $-$15,380 $\pm$ 90 & 5.00$\pm$0.39 & 2.72 $\pm$ 0.20\\
H$\gamma$ &  4341.69 &  0.044694 & $-$15,530 $\pm$ 120 & 3.55$\pm$0.86$^a$ & 2.32 $\pm$ 0.25\\
H$\delta$ &  4102.90 &  0.022105 &  $-$15,390 $\pm$ 80 & 5.16$\pm$0.13$^a$ & 2.98 $\pm$ 0.20\\
H$\epsilon$ &  3970.00 &  0.012711 &  $-$15,870 $\pm$ 120& 4.11$\pm$0.16 & 3.01 $\pm$ 0.29\\
HeI* &  3888.65 &  0.064474 &  $-$15,640 $\pm$ 160 & 2.44$\pm$0.14 & 2.32 $\pm$ 0.35
\enddata
\tablecomments{EW measured in TDE rest frame. \\
$^a$Measurements are subject to contamination from nearby spectroscopic features. }
\end{deluxetable}

\begin{deluxetable}{lcccc}
\tabletypesize{\footnotesize}
\tablewidth{0pt}
\tablecolumns{6}
\tablecaption{Detected UV absorption lines in AT2018zr\label{tab:uv_absorption}}
\tablehead{\colhead{Line} &  \colhead{$\lambda_0$} & \colhead{$\lambda_c^1$} &   \colhead{velocity} &  \colhead{FWHM} \\
 & \AA & \AA & \colhead{\kms} & \colhead{\kms}
}
\startdata
NV    &  1240        & 1182--1203$^2$  &   \nodata     & \nodata \\
SiII  &  1263        & 1182--1203$^2$  &   \nodata     & \nodata \\
OI    &  1302        & 1230            &   16,900      &   3,300 \\
CII   &  1334.43     & 1263            &   16,500      &   2,000 \\
SiIV  &  1398.0      & 1320            &   18,000      &   9,800 \\
NIII  &  1500.0      & 1422            &   16,100      &   1,800 \\
CIV   &  1548.20     & 1460            &   19,000      &   8,800 \\
AlIII &  1856.76     & 1760            &   16,500      &   1,900 \\
CIII] &  1909.0      & 1810            &   16,000      &   1,1000 \\
MgII  &  2796.3, 2803.4  & 2652        &   16,200      &   2,600
\enddata
\tablenotetext{1}{Central wavelength of the transition measured in host rest frame.}
\tablenotetext{2}{Cannot be determined due to blending with neighboring lines.}
\end{deluxetable}

The relative line intensities of multiple hydrogen Balmer transitions are
often used to constrain neutral hydrogen column density. Assuming all photons
irradiated by the continuum pass through the same amount of gas at a given
velocity, the observed radiation should have the form
\begin{equation}
I_\lambda = I_c e^{-\tau_\lambda} + B_\lambda(T_l)(1 - e^{-\tau_\lambda}) \;,
\end{equation}
where $I_c$ is the radiation of the continuum source and $T_l$ is the line
excitation temperature. Assuming that $T_l$ is negligible compared to the
continuum temperature ($T_c$), the relative line depression can be expressed
as
\begin{equation}
\label{eqn:absorption}
I(v) = \frac{I_c - I_\lambda}{I_c} = C(v)(1 - e^{-\tau(v)}),
\end{equation}
where $I(v)$ is the normalized intensity of the absorption trough,
$C(\lambda)$ is the line-of-sight (LOS) covering factor of the absorber in percentage and
$\tau(v)$ is the optical depth of the given transition. We note that in the
case where the above assumption is invalid ($T_l$ is non-negligible), the EW
of the line will be altered by a factor of $(1 - B_\lambda(T_l)/I_c)$. This
effect on the observed line EW is degenerate with that of the covering factor.

The column density $N$ of an ionic species can be expressed as
\begin{equation}\label{eqn:column}
\begin{split}
N &= \frac{m_e c}{\pi e^2 f \lambda_0} \int \tau(v) dv \\
  &= \frac{3.768\times 10^{14}~\text{cm}^{-2}}{f \lambda_0} \int \tau(v) \, dv \;,
\end{split}
\end{equation}
where $f$ is the oscillator strength and $\lambda_0$ is the rest wavelength in
\AA.
The optical depth ratio can be derived under the assumption that the absorbing
gas is in local thermodynamic equilibrium.
The relative line center opacity ($\tau$) between transitions 1 and 2 can be
simplified to (Arav et al. 2010)
%
\begin{equation}
\label{eq:boltzmann}
\frac{\tau_1}{\tau_2} = \frac{g_1 f_1 \lambda_1}{g_2 f_2 \lambda_2} \; ,
\end{equation}
where $g$ is the statistical weight of the lower level, $f$ is the oscillator
strength, and $\lambda$ is the wavelength of the transition.

It is clear from the line ratios that the H Balmer absorption lines are quite
saturated. If the H Balmer absorptions were optically thin, the EW ratio for
the H Balmer series should be close to
$\tau_{\Ha}$:$\tau_{\Hb}$:$\tau_{\Hgamma}$:$\tau_{\Hdelta}$:$\tau_{\text{H}\epsilon}$
$\approx$ 81.5:\,11.2:\,3.8:\,1.8:\,1. However, as shown in \autoref{tab:eqw},
the observed EW of all the H Balmer lines are very similar, suggesting
these lines are largely suppressed due to non-negligible optical depth.

From the normalized spectrum, we also noticed that the absorptions are
non-black, meaning they do not extend down to zero. This is a clear sign of
partial covering, where the absorbing material does not cover the
photoionizing continuum entirely.

We first model the observed Balmer absorption lines with
\autoref{eqn:absorption} by locking the relative opacity for each transitions
and assuming a constant covering factor ($C_0$) across all the hydrogen line
profiles. The best-fit parameters from directly fitting the observed spectrum
are $C_0=0.2$ and $\tau_{0,\Hb} =1.5$. We convert this to a \HI$(n = 2)$
column density of $1.5 \times 10^{15}$~\cmsq\ with \autoref{eqn:absorption}.
However, the optical depth and therefore the \HI$(n=2)$ column density derived
from direct fitting may underestimate the actual values. The opacity is
strongly suppressed by the fitting routine since the model line shape becomes
flat-topped as it saturates, while the observed absorption features are more
peak-like.
One possibility is that each absorption trough is comprised of multiple
unresolved narrow components with different velocities rather than a single
broad component.

\begin{figure}[htb]
\centering
\includegraphics[width=3.5in, angle=0]{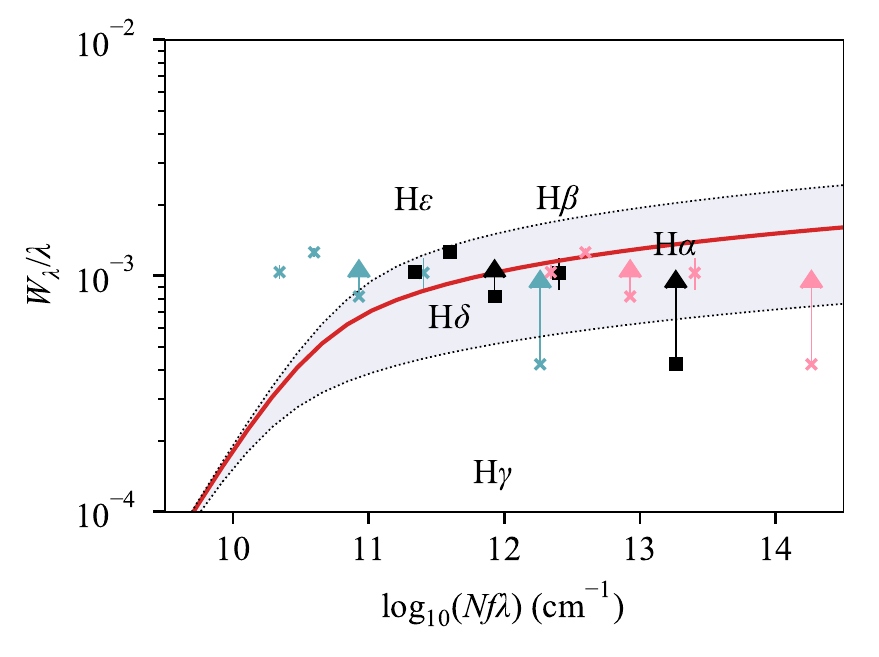}
\caption{Curve of growth analysis of the Balmer absorption lines. Only \Hb,
\Hdelta, and H$_\epsilon$ are used in the fit. The best-fit curve with $b =
313$~\kms, $C=0.25$ and $\log_{10} N = 17.6$~\cmsq\ is shown in red. The
shaded area shows the region bounded by the 1$\sigma$ uncertainty in the
Gaussian parameter ($\sigma_b = 170$~\kms). We also plot our data with
$\log_{10} N = 16.6$~\cmsq (green) and $\log_{10} N = 18.6$~\cmsq (pink) on
the growth curve. In this high opacity regime ($\tau > 1$), the equivalent
width is insensitive to $N$.}
\label{fig:cog}
\end{figure}

Rather than fitting the line profile, we instead use the curve of growth
method to model the total opacity summed over the H Balmer transitions. We
exclude the use of the \Ha\ and \Hgamma\ in the fit due to uncertain
contribution from nearby emission lines. Shown in \autoref{fig:cog}, our
best-fit Gaussian parameter ($b$), ionic column density ($N$), and covering
factor are $b = 313 \pm 170$~\kms, $\log_{10} N_{H\,\text{\sc I}(n=2)} = 17.6
\pm 0.9$~\cmsq, and $\log_{10} C = -0.6 \pm0.2$, respectively. The high
opacity of the Balmer lines place the data points on the logarithmic regime of
the curve, where the equivalent width ($W_\lambda$) is relatively insensitive to $N$
($W_\lambda \propto b\sqrt{\ln(N/b)}$). The large uncertainty in $b$ may be
attributed to the fact that this parameter is degenerate with the covering
factor. However, the absorber needs to cover at least 15\% of the continuum as
measured from \Hb. Therefore, $b$ must be narrower than $\sim$\,500~\kms,
which translates to a FWHM of $\sim$\,900~\kms. Considering these effects, the
true $N_{H\,\text{\sc I}(n=2)}$ should be even higher, placing a lower limit
for $\tau_{\Hb} \gtrsim 720$.

\subsection{UV Spectroscopy}
\label{subsec:HST_UV_data}

\begin{figure*}
\centering
\includegraphics[width=7.0in, angle=0]{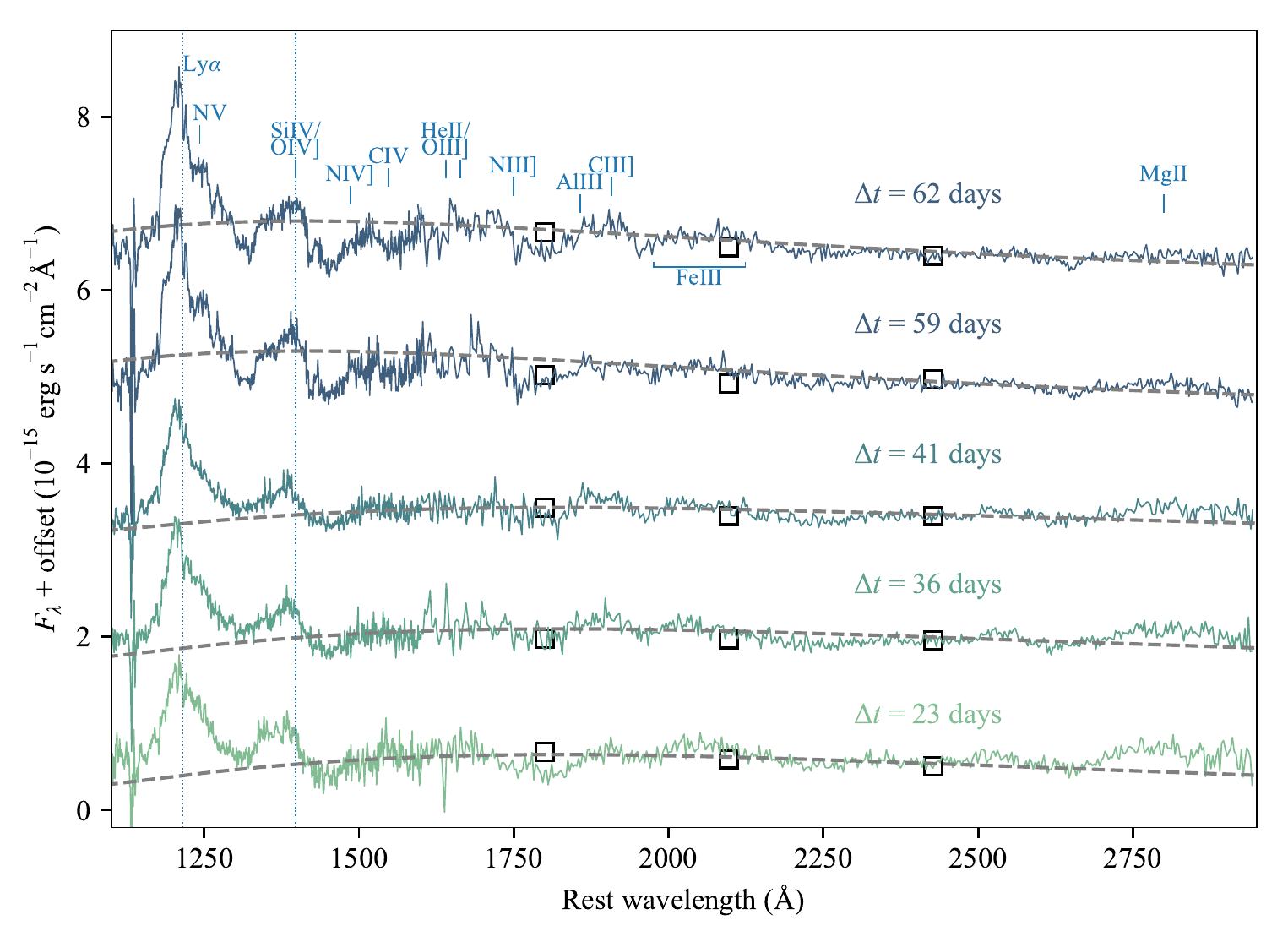}
\caption{\hst\ STIS spectra of AT2018zr observed at different epochs, where
$\Delta t$ indicated the rest-frame time elapsed since peak in the $r$ band
(MJD 58194.5). To guide the eye, the grey dashed line marks the tentative FUV
continuum extrapolated from a blackbody spectrum that is used to describe the
NUV and optical SED. The blackbody spectra correspond to a temperature of $T =
1.7 \times 10^4$~K for the first 3 epochs and $T = 2.2 \times 10^4$~K for the
last two epochs. The black squares indicate the interpolated broadband flux
measured by \swift.}
\label{fig:multi_epoch_uvspec}
\end{figure*}

We present 5 epochs of \hst\ spectra in \autoref{fig:multi_epoch_uvspec}, where
blackbody spectra with temperatures derived from the broadband NUV and optical
photometry (see \citealt{2019apJ...872..198V}) are overplotted with grey
dashed lines. Traditionally, observers found the UV and optical photometry
measured by \swift\ in TDEs in agreement with a blackbody spectrum with $T
\sim \text{few} \times 10^4$~K \citep[e.g.][]{2016MNRAS.463.3813H,2017ApJ...842...29H}. However,
this agreement has not been carefully examined blueward of the \swift\ $uvw2$
band (central wavelength 1928~\AA). In fact, previous works have also
suggested that a significant amount of TDE radiation may be emitted in the
extreme UV \citep[e.g.][]{2018ApJ...859L..20D}. Therefore, while the blackbody
spectrum captures the general shape of the NUV spectra in all 5 epochs, the
FUV continuum cannot be determined as accurately. We also note that this
blackbody spectrum cannot account for the X-ray flux that was observed in
AT2018zr. A second blackbody of $T \sim 100$~eV is required to describe the
entire SED \citep{2019apJ...872..198V}.

The UV spectra of AT2018zr show complex features, where broad emission and
absorption are variable and often blended together, making it difficult to
measure these lines accurately. In particular, the high-velocity BAL features
(\autoref{subsec:absorptions}) only become discernible in the UV spectra in
later epochs. Despite the uncertainties in FUV continuum placement, we conclude
the evolution of the FUV lines by assuming the underlying continuum is a single blackbody, with
temperatures extrapolated from NUV and optical observations. Here we
summarize the observed properties of these features qualitatively.

\begin{enumerate}
    \item The \La\ emission line is blueshifted by about $3,\!000$~\kms\ in
    all 5 epochs. Assuming no additional source of FUV continuum flux, \La\
    becomes $\sim$50\% more prominent after $\Delta t \approx 59$ days.

    \item High ionization \NV\ emission may be present but is blended with the
    red wing of the broad \La,  therefore the \NV\ emission peak cannot be
    individually resolved.

    \item High ionization emission line \SiIV\ is also blueshifted with
    respect to the rest frame of the host.

    \item In the first two epochs, there is a marginally detected broad
    feature (FWHM $\sim 1.3 \times 10^4$~\kms) at the rest wavelength of
    \MgII.

    \item We do not observe an enhancement in the N/C ratio as reported in previous TDEs ASASSN-14li, iPTF16fnl, and iPTF15af \citep{Cenko2016,2018MNRAS.473.1130B,2018arXiv180907446B}.

    \item None of the emission peaks shifted significantly over the monitoring
    period, which suggests that the kinematics of the UV line emitting region
    did not vary much in the first two months.

    \item The absorption features at $v\sim15,\!500$~\kms\ are weak or
    completely absent in the first 3 epochs. In the last two epochs,
    absorption lines including high ionization lines \NV, \SiIV, \CIV, and low
    ionization lines \OI, \CII, \SiII, \CIII, \AlIII, and \MgII\ are detected
    in the spectra. These UV absorption features detected in AT2018zr fit in
    the low-ionization BAL (LoBAL) category.

    \item The UV broad absorption lines are saturated and non-black,
    suggesting partial covering of the continuum source. In the later two
    epochs, the UV absorptions are also seen to be shallower than the emission
    lines (e.g. \La), which indicates that the BAL system does not cover the
    line emitting region entirely, either.
\end{enumerate}

\subsection{Optical Spectroscopy}

\begin{figure*}
\centering
\includegraphics[height=8.3in,angle=0]{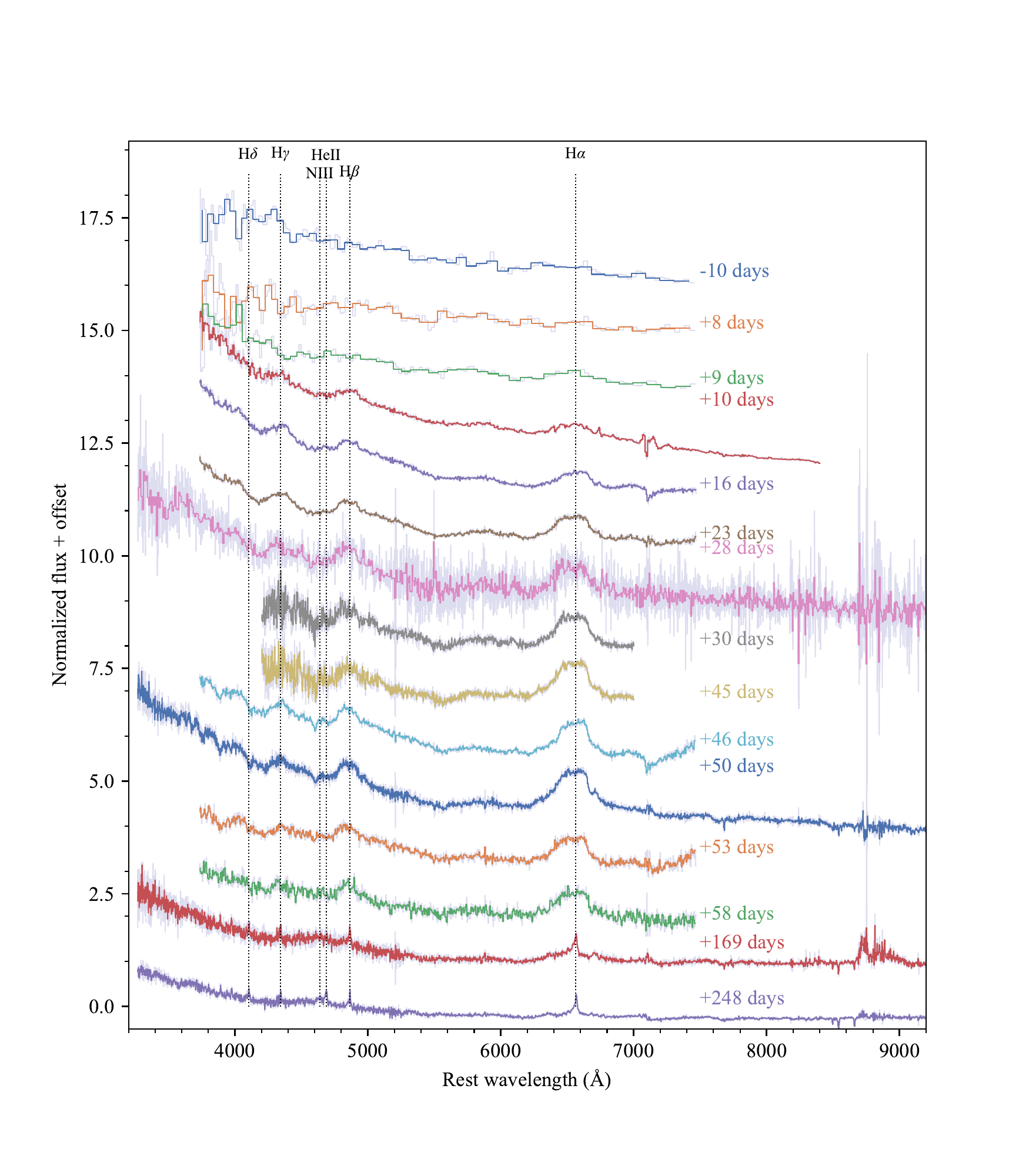}
\caption{Host-subtracted optical spectra of AT2018zr color-coded by epoch. The
first 3 spectra (from top) were obtained by the low-resolution SEDm. The
telescope and instrument that took each spectrum are listed in
\autoref{tab:obs_spec}. The original data are plotted in pale grey under the
smoothed spectra. All spectra shown here have been binned in wavelength by 3
except for the Keck spectrum on day 31, which is binned by 5 pixels due to
higher noise. The corresponding phase ($\Delta t$) is indicated to the right
of each spectrum.
\label{fig:nd_all_spec}}
\end{figure*}

At optical wavelengths, host galaxy contributes a non-negligible amount of
flux to the observed spectrum. Since there was no pre-flare spectrum of the
host galaxy, we estimated the host flux by fitting a synthetic galaxy spectrum
with SDSS model magnitudes \citep{2019apJ...872..198V}. To perform subtraction
of the host flux, we calibrated the flux level in each optical spectrum
against \swift\ V-band photometry, interpolated to the spectroscopic epochs.
We then convolved the synthetic host galaxy spectrum with a Gaussian kernel to
account for instrumental broadening and subtracted the broadened, synthetic
spectrum from our observed spectra. A montage of the host-subtracted spectra
is shown in \autoref{fig:nd_all_spec}. The flux is normalized to the
5500--6000~\AA\ region in rest wavelength and offset from each other for
better visualization.

We describe our analysis of the optical spectra before and after the
Sun-constrained break separately, because the line profiles were dramatically
different after the break.

Both broad \Ha\ and \Hb\ lines were readily detected in the spectra from day
10 onward. The strength of these lines grew monotonically with time in the first two months.
In the earlier set of spectra with good S/N, we noticed that the \Ha\ and \Hb\
emission profiles were asymmetric, with what appeared to be a `dent' in the
red wing (\autoref{fig:normalized}). Although the positions of these `dents'
are consistent with He\,{\sc i}\,$\lambda7065$ and Fe\,{\sc ii} blueshifted by
the same velocity as the BAL system, the width and depth are broader and
stronger than the other optical absorption lines. Given the similarity in the
line profile shapes of \Ha\ and \Hb\ (\autoref{fig:gaussian_fit} and
\autoref{fig:late_time_optical}), this asymmetry is more likely the result of
emission line region geometry than contamination from neighboring absorption
lines.

Interestingly, the flat-topped H Balmer emission profiles in AT2018zr in the
earlier monitoring period differ from the Gaussian emission lines seen in
other TDEs. This motivated \cite{2018arXiv180802890H} to fit these lines with
a model that combines the effects of an elliptical disk, spiral arm, and wind.
Here, we use two different approaches to model the observed lines, including
(1) phenomenological fitting with two Gaussians and (2) radiative transfer
equation in a spherical outflow.

\subsubsection{Double Gaussian Model}

\begin{figure}[htb]
\centering
\includegraphics[width=3.5in, angle=0]{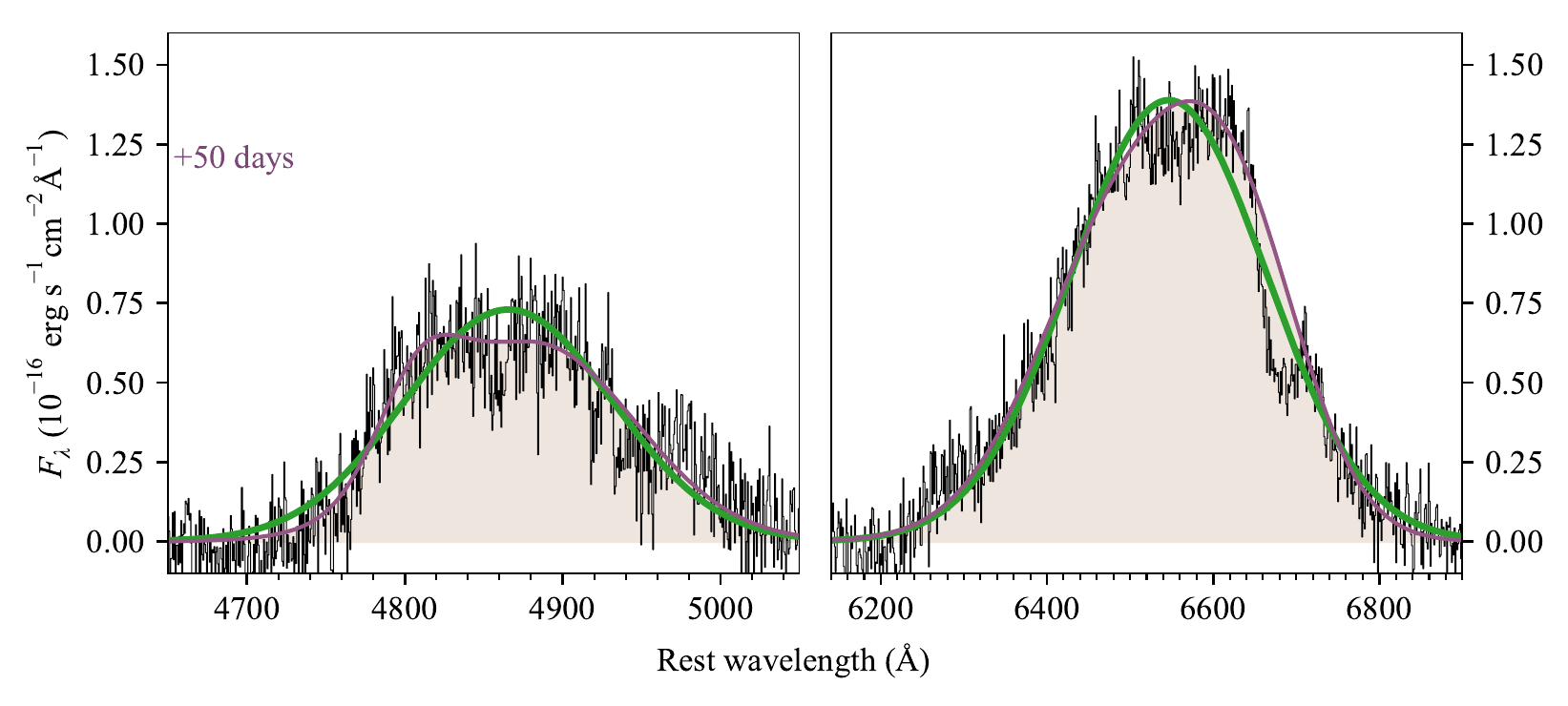}
\caption{Model fit to \Ha\ (right) and \Hb\ (left) lines in the Keck LRIS
spectrum from 2018 May 10. We show the best-fit result of a single Gaussian
(green) and a double Gaussian model (purple) with solid line. It is clear from
the plot that the flat-topped profile that cannot be fitted well by a single
Gaussian.}
\label{fig:gaussian_fit}
\end{figure}

In our initial attempt to fit these lines, we noticed that the line profiles,
especially \Ha, cannot be well described by a single Gaussian (green solid
line in \autoref{fig:gaussian_fit}). Therefore, we chose to fit each
flat-topped Balmer line with two Gaussians simultaneously. Note that the
choice of model was not motivated by the underlying physics of line formation,
but rather to obtain satisfactory description of the line profiles. This
allows us to measure the FWHM and line luminosity for comparison with other
TDEs.

We first measured the continuum level in the spectral lines by performing a
quadratic fit to the region enclosing the lines (4630--5150~\AA\ for \Hb\ and
5950--6950~\AA\ for \Ha). Although we have subtracted the synthetic galaxy
from our observed spectra prior to measuring the lines, we find the spectra
near \Ha\ and \Hb\ line centers to be noisy, and may still have residual
contribution from the host. Therefore, we masked the line centers while
fitting the emission lines. In the two-component fit, we set constraints for
the two components to center blueward and redward of the rest wavelength of
the line and allowed the linewidths and amplitudes to vary freely.

The best-fit two-component model is marked in purple in
\autoref{fig:gaussian_fit}. From the best-fit models, we measured the
luminosity, equivalent width (EW), and FWHM for both \Ha\ and \Hb\ lines.
These parameters are tabulated in \autoref{tab:param_opt_spec_ha} and
\autoref{tab:param_opt_spec_hb}.

\begin{deluxetable}{lccc}
\tabletypesize{\footnotesize}
\tablewidth{0pt}
\tablecolumns{4}
\tablecaption{\Ha\ line parameters.\label{tab:param_opt_spec_ha}}
\tablehead{
\colhead{Phase} & \colhead{FWHM} & \colhead{EW}& \colhead{Luminosity}  \\
\colhead{(days)} & \colhead{($10^4$~\kms)} & \colhead{(\AA)} & \colhead{($10^{41}$ erg s$^{-1}$)}
}
\startdata
10 & 1.35$\pm$0.81 & 86$\pm$17& 2.0$\pm$0.2\\
16 & 1.26$\pm$0.17 & 89$\pm$4& 1.9$\pm$0.3\\
23 & 1.39$\pm$0.12 & 146$\pm$4& 3.4$\pm$0.4\\
28 & 1.42$\pm$1.62 & 188$\pm$403& 4.0$\pm$8.9\\
30 & 1.42$\pm$0.10 & 198$\pm$3& 4.4$\pm$0.4\\
45 & 1.39$\pm$0.09 & 224$\pm$2& 5.2$\pm$0.2\\
46 & 1.28$\pm$0.13 & 202$\pm$7& 3.8$\pm$0.4\\
50 & 1.44$\pm$0.21 & 265$\pm$26& 5.3$\pm$1.5\\
53 & 1.35$\pm$0.13 & 159$\pm$4& 3.2$\pm$0.3\\
58 & 1.42$\pm$0.38 & 190$\pm$11& 4.2$\pm$0.3\\
169 & 0.43$\pm$0.18 & \nodata & 0.2$\pm$0.1\\
248 & 0.35$\pm$0.07 & \nodata   & $<$0.1\\
169$^n$ & 0.11$\pm$0.07 & \nodata & (8.7$\pm$2.0)$\times10^{-2}$\\
248$^n$ & 0.07$\pm$0.03 & \nodata    & (4.9$\pm$0.8)$\times10^{-2}$
\enddata
\tablecomments{$^n$ denotes the measurements for the narrow components. We do not calculate the equivalent width in late time spectra due to the complexity in separating the broad and narrow components.}
\end{deluxetable}

\begin{deluxetable}{lccc}
\tabletypesize{\footnotesize}
\tablewidth{0pt}
\tablecolumns{4}
\tablecaption{\Hb\ line parameters.\label{tab:param_opt_spec_hb}}
\tablehead{
\colhead{Phase} & \colhead{FWHM} & \colhead{EW}& \colhead{Luminosity}  \\
\colhead{(days)} & \colhead{($10^4$ \kms)} & \colhead{(\AA)} & \colhead{($10^{41}$ erg s$^{-1}$)}
}
\startdata
10 & 1.14$\pm$0.22 & 21$\pm$2& 1.0$\pm$0.2\\
16 & 1.05$\pm$0.56 & 27$\pm$11& 1.1$\pm$0.6\\
23 & 1.23$\pm$0.18 & 33$\pm$3 & 1.2$\pm$0.2\\
28 & 1.17$\pm$2.78 & 46$\pm$283& 1.6$\pm$5.3\\
30 & 1.17$\pm$0.22 & 42$\pm$2& 1.4$\pm$0.2\\
45 & 0.96$\pm$0.63 & 39$\pm$13& 1.2$\pm$0.5\\
46 & 0.96$\pm$0.59 & 41$\pm$11& 1.3$\pm$0.5\\
50 & 1.20$\pm$0.09 & 47$\pm$1& 1.5$\pm$0.1\\
53 & 1.11$\pm$0.36 & 27$\pm$12& 0.9$\pm$0.4\\
58 & 0.89$\pm$1.04 & 34$\pm$48& 1.1$\pm$0.9\\
169 & 0.15$\pm$0.05 &  \nodata & $<$0.1 \\
248 & 0.19$\pm$0.02 &  \nodata  & $<$0.1 \\
169$^n$ & 0.05$\pm$0.02 & \nodata & (2.0$\pm$1.4)$\times10^{-2}$ \\
248$^n$ & 0.04$\pm$0.01 & \nodata & (1.8$\pm$0.4)$\times10^{-2}$
\enddata

\end{deluxetable}

\subsubsection{Spherical Outflow Model for Emission Lines}

\cite{2018ApJ...855...54R} demonstrated that in TDE outflows, the blue absorption wing of
a classical P-Cygni profile may appear mostly as emission if the line excitation
temperature is sufficiently high. Given the similarity of their theoretical
line profile to our observations, we also fit the \Ha\ emission with this
radiative transfer model proposed by \cite{2018ApJ...855...54R} for three
different epochs.

We consider a spherically symmetric, homologously expanding ($v \propto r$)
medium. A continuum photosphere is located at radius $r_\text{ph}$, which is
responsible for emitting the observed continuum flux at wavelengths near \Ha\
at the time of interest. The gas density falls off as $r^{-2}$ at radii beyond
$r_\text{ph}$ out to a maximum radius $r_\text{out}$, where the velocity is
$v_\text{max}$. The line absorption opacity $\kappa_l$ and the line Doppler
width $v_D$ are set to single values at all positions outside the continuum
photosphere. The line excitation is set by $T_\text{ex} = a + b/r$ where $a$
and $b$ are chosen so that the line excitation temperature equals a specified
value $T_\text{ex,ph}$ at the photosphere, and $T_\text{ex,out}$ at
$r_\text{out}$. The Sobolev approximation can then be used to calculate the
observed emission by integrating the line source function along lines of sight
passing through the line-emitting material, as in \citet{Jeffery1990,
2018ApJ...855...54R}. The strength of the emission line with respect to the
continuum flux is then related to the Sobolev depth $\tau_S$ at each radius:
\begin{equation}
\tau_{S}(r) = \sqrt{\pi} \, \rho(r) \, \kappa_l v_D  \left(
\frac{dv}{dr}\right)^{-1} \,\, .
\end{equation}
For the conditions considered here, the electron scattering optical depth is
negligible for the line photons.

\begin{figure*}[htb]
\centering
\includegraphics[width=\textwidth, angle=0]{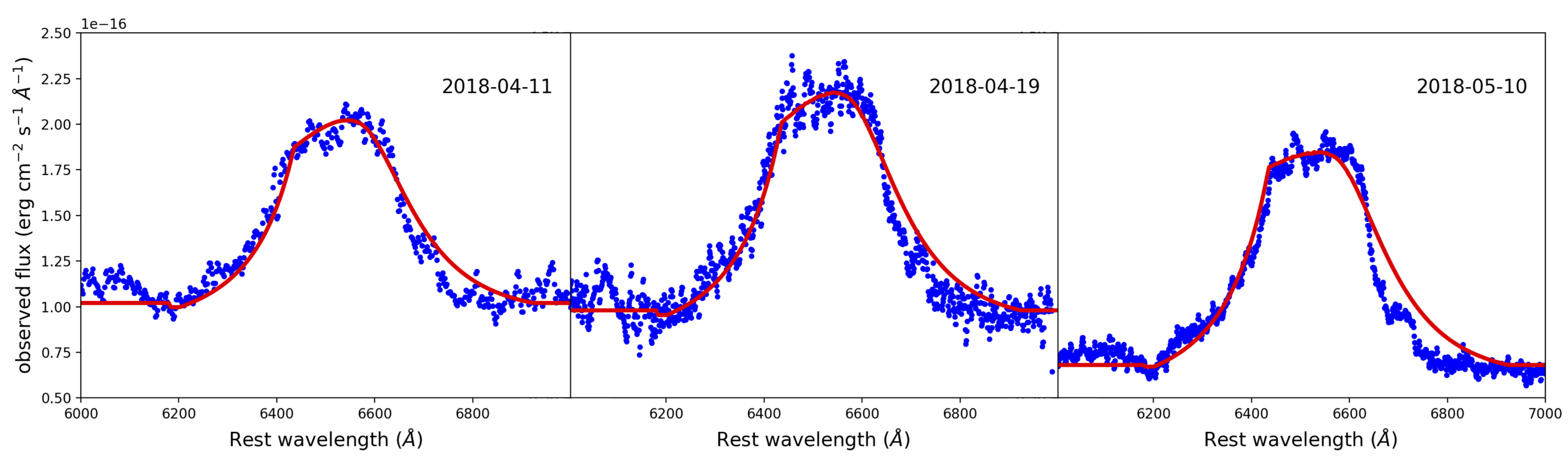}
\caption{\Ha\ line profiles, with fits to a spherical outflow model for the broad emission. An additional absorption component at the highest velocities is discussed in \autoref{subsec:absorptions}.}
\label{fig:combined_Halpha_line_profiles}
\end{figure*}

Rather than perform a full multi-dimensional fit, we set all the parameters to
the fiducial values listed in \autoref{tab:halpha_param} and then adjusted the value of
$\tau_S(r_\text{ph})$ between spectral epochs until we judge by eye that a
satisfactory fit has been achieved. The continuum flux is also adjusted
between epochs to match what is observed. The resulting values are listed in
\autoref{tab:halpha_opacities}. Given the definition of $\tau_S$, the values
of $\kappa_l$ and $v_D$ are degenerate in this formulation, since only their
product enters into the calculation, as long as $v_D \ll v_{\rm max}$. The results are plotted in \autoref{fig:combined_Halpha_line_profiles}

\begin{deluxetable}{lccccc}
\tabletypesize{\footnotesize}
\tablewidth{0pt}
\tablecolumns{6}
\tablecaption{Fiducial parameters for the spherical outflow model of the \Ha\
line profiles. These values were kept the same between the 3 spectral epoch
fits. The gas density is assumed proportional to $r^{-2}$, and the gas
velocity is assumed proportional to $r$ (and is entirely radially outflowing).
The line excitation temperature drops as $a + b/r$ where $a$ and $b$ are
chosen to match the listed excitation temperature values at the photosphere
radius and the outer radius.\label{tab:halpha_param}}
\tablehead{
\colhead{$r_{\rm ph}$} &  \colhead{$r_{\rm out}$} & \colhead{$v_{\rm max}$} & \colhead{$\rho_{\rm ph}$} & \colhead{$T_{\rm ex,ph}$}  & \colhead{$T_{\rm ex,out}$} \\
\colhead{(cm)} & \colhead{(cm)} & \colhead{(10$^4$ \kms)} & \colhead{(g cm$^{-3}$)} & \colhead{(K)} & \colhead{(K)}
}
\startdata
    $10^{15}$ & $3 \times 10^{15}$& 1.75 & $1.16 \times 10^{-15}$ & $3 \times 10^4$ & $10^4$
\enddata
\end{deluxetable}

\begin{deluxetable}{lccc}
\tabletypesize{\footnotesize}
\tablewidth{0pt}
\tablecolumns{4}
\tablecaption{Continuum fluxes and Sobolev depths used in the \Ha\ line
profile fits, which were varied between spectral
epochs.\label{tab:halpha_opacities}}
\tablehead{
\colhead{$\Delta t$ (days)} &
\colhead{23} &  \colhead{30} & \colhead{50}
}
\startdata
    $\tau_S(r_\text{ph})$ & 0.39 & 0.48 & 0.47 \\
    Continuum flux & $1.02 \times 10^{-16}$  & $9.80 \times 10^{-17} $ &$6.81 \times 10^{-17}$ \\
    (erg cm$^{-2}$ s$^{-1}$ \AA$^{-1}$)  & & & \\
\enddata
\end{deluxetable}

\subsubsection{Late-Time Optical Spectrum}
\label{subsubsec:late_time_Keck}

\begin{figure}
\centering
\includegraphics[width=3.0in, angle=0]{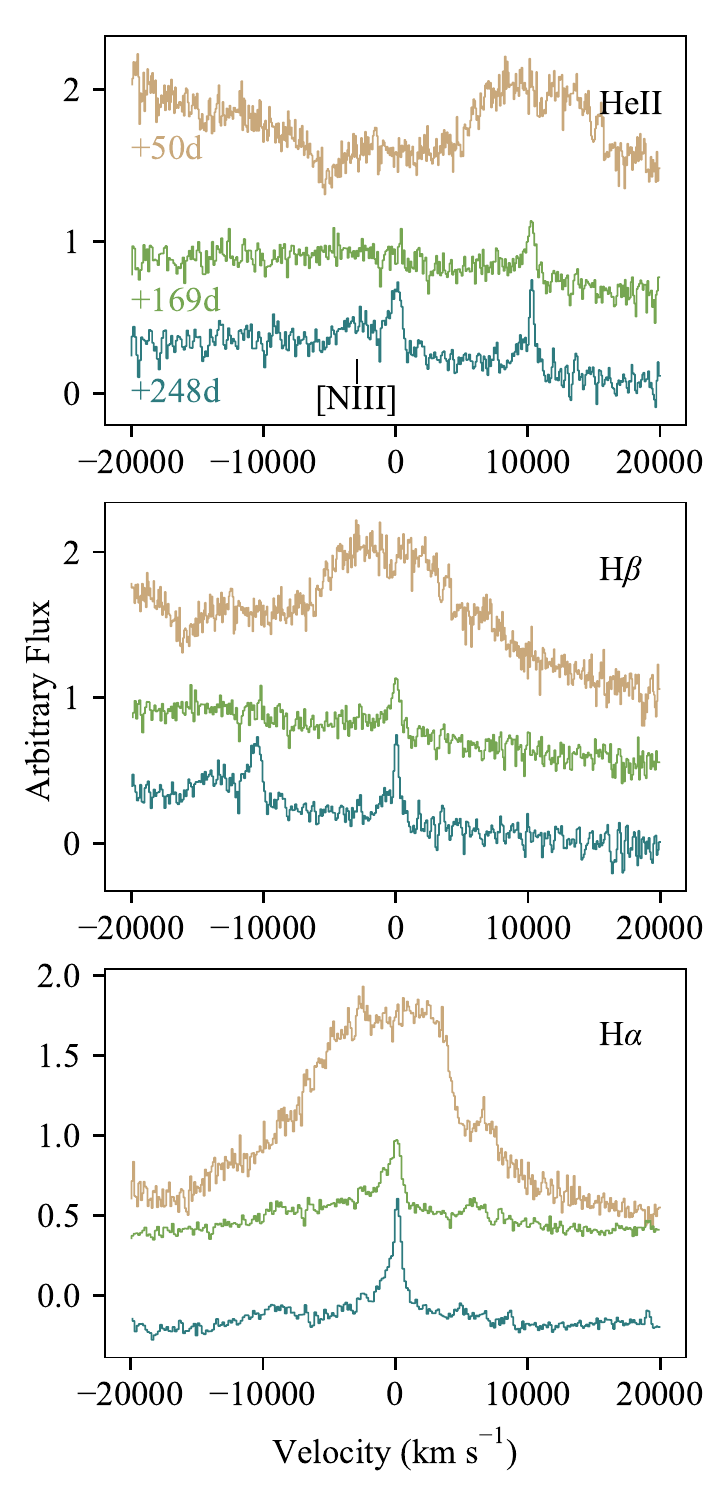}
\caption{Velocity profiles of \HeII\, \Hb\, and \Ha\ of AT2018zr at $\Delta t
= 50$~d, $\Delta t = 169$~d, and $\Delta t = 248$~d. Narrow \Ha, \Hb, and
\HeII\ emission have emerged 6 months after maximum light. The narrow line
widths are on the order of $1,000$\kms\, which is only $\sim$10\% of the broad
Balmer line widths from earlier epochs. We note that on day 50, \Ha\ and \Hb\
have a similar velocity profile that is dented on the red wing, which is
likely to arise from the geometry of the line-forming region. }
\label{fig:late_time_optical}
\end{figure}

The spectroscopic features showed
dramatic changes (\autoref{fig:late_time_optical}) at late times ($\Delta t$= 169 d and 248 d), . First, \HeII\ and N\,{\sc
iii}\,$\lambda$4640~\AA, a strong transition of the Bowen fluorescence (BF)
mechanism, emission lines are readily detected on day 248, while
the detections are somewhat tentitive on day 169. In BF, the He\,{\sc
ii}\ \La\ photons at 303.783~\AA\ can excite a number of nearby transitions, which decay into optical emission lines such as O\,{\sc iii}\ (at 3133, 3429, and 3444~\AA) and N\,{\sc iii}\,$\lambda$4640~\AA. BF emission lines have been detected in systems such as planetary nebulae and X-ray binaries and also in TDE iPTF-15af \citep{2018arXiv180907446B}.

The late appearance of the \HeII\ emission is unlike most of the other TDEs
(e.g. PS1-10jh, \citealt{Gezari2012}), where hydrogen emission is often
suppressed relative to helium at early times \citep{Roth2016}. From the point
of view of photoionization, the absence of the
\HeII\ line may be attributed to the lack of He\,{\sc ii} ionizing photons
($\chi_\text{ion} = 54.4$~eV) at earlier times. This is supported by the fact
that the blackbody temperature of AT2018zr is cooler than many other TDEs at
first, but warms up drastically at later times \citep{2019apJ...872..198V}.
The hot continuum indicated by the late-time \swift\ observations would
naturally explain the emergence of the \HeII\ emission.

The broad \Ha\ and \Hb\ emission became extremely weak when we revisited this
source on $\Delta t=169$ d. Instead, narrow
\Ha\ and \Hb\ emission emerged.
The most striking difference is found in the \Ha\ and \Hb\ line, which are now
dominated by two strong narrow components whereas the broad \Ha\ and \Hb\
emission have vastly diminished, though they can still be seen. We measured
\Ha\ and \Hb\ in the late-time spectra with a double Gaussian model, where
both broad and narrow components are initially centered at the rest wavelength
of the transition. We show the best-fit results in
\autoref{tab:param_opt_spec_ha} and \autoref{tab:param_opt_spec_hb}. The newly
formed narrow \Ha\ and \Hb\ emission have FWHMs on the order of
$1,\!000$~\kms, corresponding to a virial radius of $\sim$\,$6 \times
10^{17}$~cm.

\section{Discussion}
\label{sec:discussion}

\subsection{UV Spectra of TDEs}

\begin{figure*}[htb]
\centering
\includegraphics[width=7.0in, angle=0]{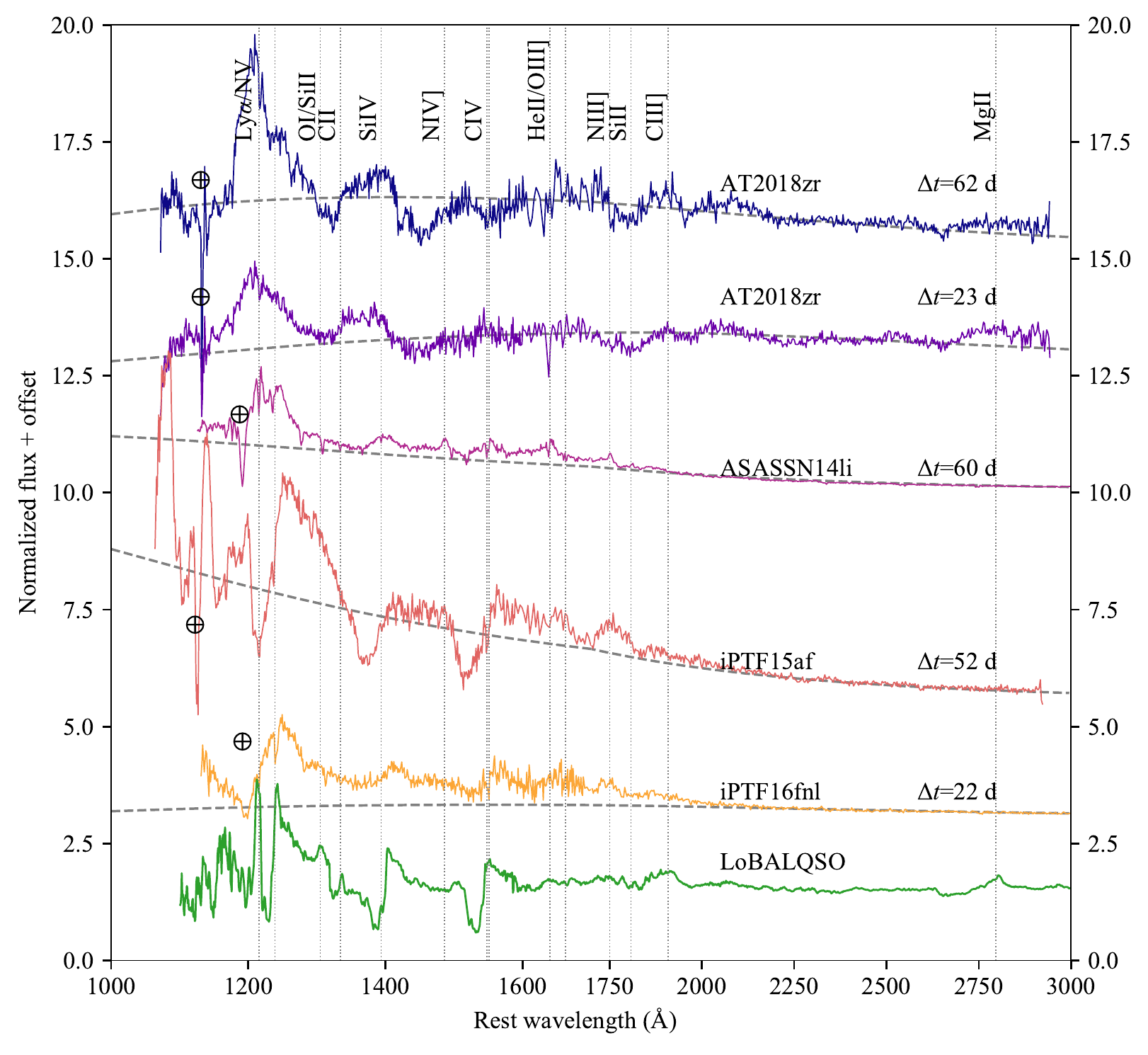}
\caption{Comparison with the UV spectra of other TDEs. The grey dashed line at
the bottom of each TDE spectrum marks the continuum, which is approximated by
a blackbody spectrum of the corresponding temperature.}
\label{fig:UVTDE_compare}
\end{figure*}

AT2018zr is the fourth optically discovered TDE with \hst\ UV spectroscopic
observations. We compare its \hst\ spectrum with the other three---ASASSN-14li,
iPTF15af, and iPTF16fnl---in \autoref{fig:UVTDE_compare} \citep{Cenko2016,2018MNRAS.473.1130B,2018arXiv180907446B}. In particular, iPTF16fnl is the only other TDE with a UV spectroscopic sequence \citep{2018MNRAS.473.1130B}. For comparison, we
also show a composite spectrum of low-ionization broad absorption line (LoBAL)
QSOs from the FIRST Bright Quasar Survey \citep{2001ApJ...546..775B} in green.

Because the high-velocity absorption features were weak or completely absent
in the first two \hst\ epochs, we are able to identify that the emission peaks
of \La\ and \SiIV\ are blueshifted with respect to the TDE rest frame.
AT2018zr is the only object that exhibit blueshifted UV emission (\La, \SiIV)
among the TDEs. In ASASSN-14li, the emission lines are near the systemic host
velocity \citep{Cenko2016} while in iPTF15af and iPTF16fnl the emission lines
are systematically redshifted \citep{2018arXiv180907446B,
2018MNRAS.473.1130B}. The velocity of the emission lines may arise from a
combination of effects, including the geometry, kinematics, and optical depth.
In iPTF15af, the redshifted broad emission lines are most likely caused by the
absorption of the blue wing at $v \approx -5,\!000$~\kms.
\cite{2018MNRAS.473.1130B} obtained 3 epochs of \hst\ spectra of iPTF16fnl,
spanning a time coverage of $\sim$ a month. While analyzing the evolution of
iPTF16fnl, they found different evolution of the central wavelength for high-
and low-ionization lines. Specifically, they found the high-ionization
emission lines were initially redshifted by $v \sim 2,\!000$~\kms\ but then
evolved to peak near the wavelengths of the corresponding transitions at later
times while the low-ionization lines showed no apparent shift at any given
time. Since the UV spectrum of iPTF16fnl exhibits blueshifted absorption troughs for \SiIV\ and \CIV\ (FWHM$\sim$6000\kms), they suggest that the redshifted emission peaks may be due to the blue wings of the emission lines being significantly absorbed, as a result of an outflow, at
earlier times. This would be consistent with our finding that the BAL system
is fast-evolving, on a timescale of days to weeks.

\begin{figure}
\centering
\includegraphics[width=3.5in, angle=0]{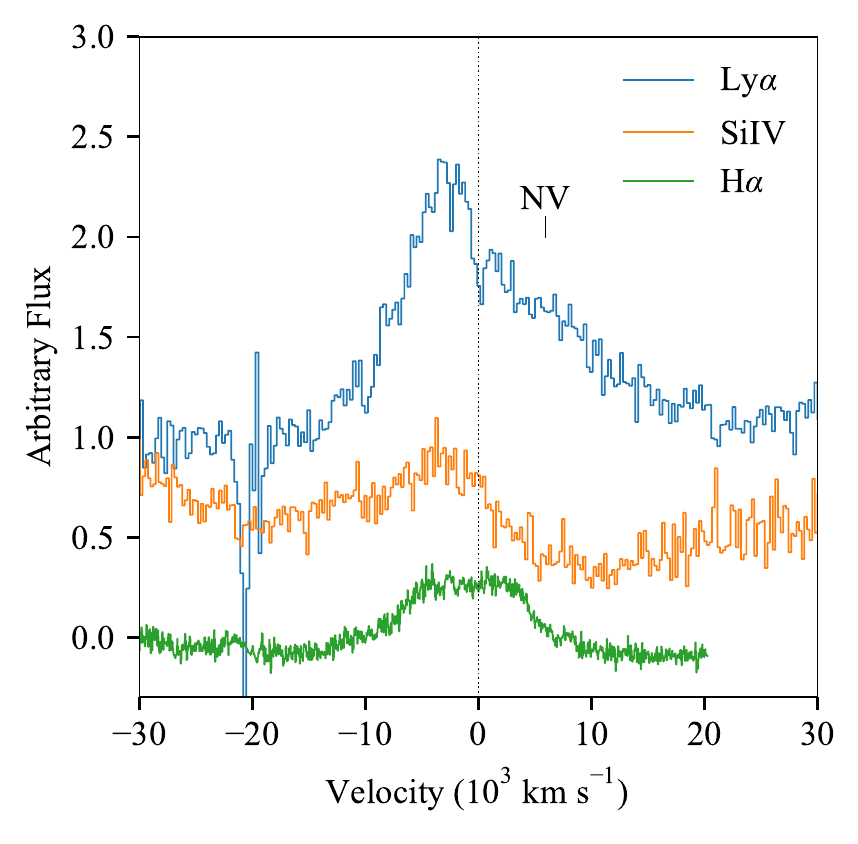}
\caption{The velocity profiles of the \La, \SiIV, and \Ha\ emission lines at
$\Delta t = 36$~d. Both the \La\ and \SiIV\ show a systemic blueshift of
$\sim$\,3,000~\kms\ relative to the \Ha, which suggests the lines were formed
in locations with different kinematics.}
\label{fig:line_vel}
\end{figure}

It is commonly observed in QSOs that the high-ionization lines (e.g.\ \CIV)
are systematically blueshifted with respect to the low-ionization lines
\citep[e.g.][]{1982ApJ...263...79G, 2002AJ....124....1R}. This blueshift is
often attributed to the presence of a radiatively driven wind
\citep{1995ApJ...451..498M}, which may also be responsible for blueshifting
($v \approx 3,\!000$~\kms) the \SiIV\ and even the \La\ emission in AT2018zr.
In fact, the velocity profile of the strong, blueshifted \La\ emission is more
similar to the \SiIV\ emission than to the \Ha\ emission
(\autoref{fig:line_vel}).
In the class of AGNs with double-peaked broad Balmer emission lines, it has
been observed that the \La\ emission tends to be single-peaked and narrower in
width \citep{2003ApJ...599..886E}. This difference has been attributed to \La\
being emitted from a higher-ionization gas originating in a wind, compared to
higher-density gas in the accretion disk producing the low-ionization lines
\citep{2003ApJ...599..886E}.
Based on the similar velocity, the observed \La\ emission may naturally form
in the same outflowing gas with intermidiate velocity ($v \approx
3,\!000$~\kms) as the \SiIV\ emission.
We measured a \La\ luminosity of $5.08 \times 10^{43}$~erg s$^{-1}$ from the
first \hst\ epoch, which is at least two orders of magnitude stronger than the
\Ha\ emission observed on the same night and $\gtrsim$\,10 times larger than
the expected value from case B recombination ($\approx$\,8.7).

\subsection{Broad Balmer Emission Lines}

\begin{figure}[htb]
\centering
\includegraphics[width=3.5in, angle=0]{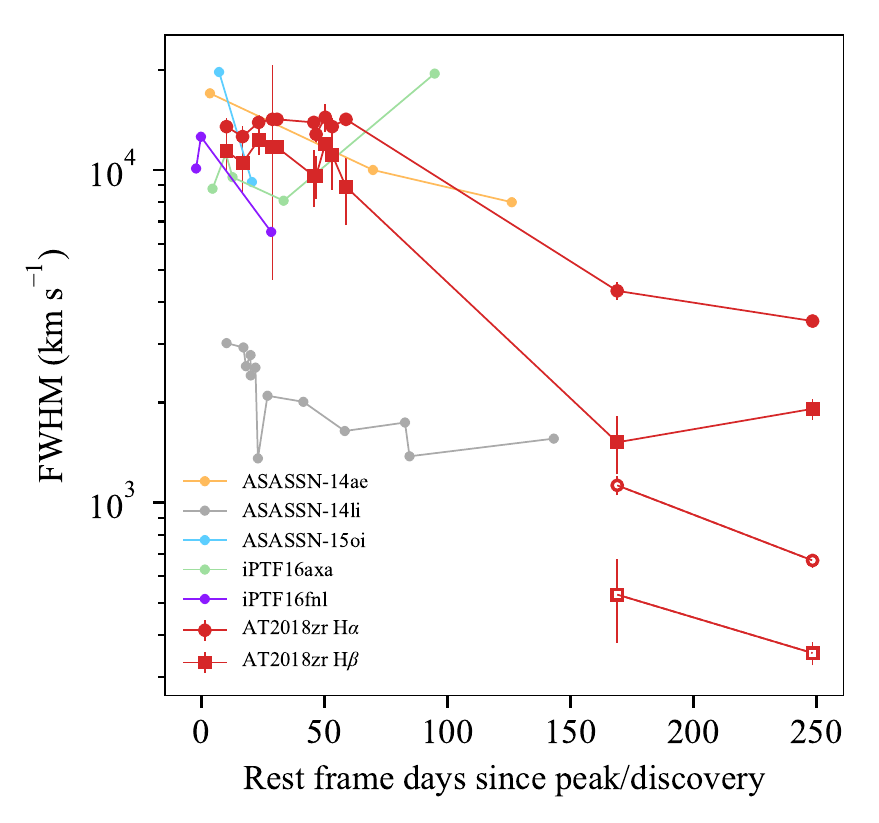}
\caption{Evolution of the FWHM of \Ha\ and \Hb\ for AT2018zr compared with the
FWHM evolution of \Ha\ of other well-studied TDEs. The \Ha\ and \Hb\
linewidths declined much slower at early times than the other TDEs. The
late-time linewidths are almost as narrow as ASASSN-14li. The FWHMs of the
narrow components are shown with open symbols.}
\label{fig:FWHM_time}
\end{figure}

In \autoref{fig:FWHM_time}, we show the evolution of the full width at
half-maximum (FWHM) of \Ha\ and \Hb\ measured with the best-fit
double-Gaussian model. The \Ha\ linewidth of AT2018zr is comparable to
ASASSN-14ae at earlier times but is broader than that of the other TDEs by a
factor of $\sim$\,30\%. The FWHMs of the broad \Ha\ and \Hb\ components show
little evolution before the 3-month-long observational gap but decreased by a
factor of $\gtrsim$\,3 at late time. This long term trend of line narrowing is
also seen in other TDEs (\autoref{fig:FWHM_time}), such as in ASASSN-14ae and ASASSN-14li \citep{Holoien2014,Holoien2016}, though the blackbody luminosity in AT2018zr did not
decrease monotonically like in the other TDEs \cite{2019apJ...872..198V}. As derived by
\cite{2019apJ...872..198V}, the blackbody luminosity of AT2018zr decreased
with time until $\Delta t_\text{peak} \sim$\,70 days post peak and remained
nearly constant up to $\Delta t_\text{peak} \approx$\,250 d. Nevertheless,
neither the line width evolution in AT2018zr nor that in other TDEs conform
with the results of AGN reverberation mapping, where the \Ha\ line width
increases as the luminosity decreases due to less recombination at outer radii
\citep{Holoien2016}, hence the absence of the low velocity dispersion components.

The earlier ($\Delta t \leq 58$) broad \Ha\ FWHM corresponds to a virial
radius of $\approx 1.5 \times 10^{15}$~cm, assuming a black hole mass of
$10^{6.9}$~\Msun\ \citep{2019apJ...872..198V}. During this period, the broad
\Ha/\Hb\ line ratio follows the case B value closely, with a mean of 3.1 and a
standard deviation of 0.8. At late time, the average \Ha/\Hb\ line ratio is
7.1$\pm$0.9. The measured narrow He\,{\sc ii}/\Ha\ ratio from the last
spectrum ($\Delta t=248$~d) is $\approx$\,0.4, which is consistent with the
case B prediction assuming solar abundance \citep{2017ApJ...842...29H}.

The flat-topped line profile in AT2018zr is unique among the optical TDEs
discovered so far. In AGNs, the flat-topped line shape is often attributed to
the orbital motion of a Keplerian disk. Double-peaked emission lines have been
observed in many AGNs, which are believed to originate from the outer part of
an accretion disk at $\approx 1000\,r_g$
\citep[e.g.][]{1989ApJ...339..742C,1994ApJS...90....1E,2003AJ....126.1720S}
illuminated by a central ionizing source. In addition to Doppler broadening,
relativistic effects are incorporated to model the asymmetries seen in the
double-peaked emission lines. For example, a circular relativistic disk model
\citep{1989ApJ...339..742C} is often employed to explain AGN spectra where the
emission lines have a double-peaked shape with a stronger blue peak. The
opposite case in which the red peak is stronger could be achieved with a
elliptical disk \citep{1995ApJ...438..610E}. Such model has been employed in
fitting TDE lines in several cases, for example, PTF09djl
\citep{2017MNRAS.472L..99L} and ASASSN-14li \citep{2018MNRAS.480.2929C}.

Motivated by the theoretical line profiles of an elliptical disk,
\cite{2018arXiv180802890H} have modeled the \Ha\ line of AT2018zr (PS18kh) at
different epochs with the combining effects of an elliptical disk, spiral arm,
and wind. Their model provides a reasonable fit to the observed \Ha\ emission
line shape.
However, we find very little evidence for the double-peaked \Ha\ line profile
as claimed by \cite{2018arXiv180802890H} in our host-subtracted spectra. We
suspect the dip around line center in their data is most likely due to the
host starlight, which may not have been cleanly removed. As shown by their
fit, the observed line center flux always drops more steeply than what the
elliptical disk model can reproduce. Furthermore, for a pure Keplerian disk, a
flat-top line is only expected under very specific conditions, i.e.\ when the
disk rotation axis is $\lesssim$\,15$\arcdeg$ \citep{2014MNRAS.439.1051L}.

In addition, the X-ray observations favor the presence of an accretion disk
\citep{2019apJ...872..198V}, which is not expected in the elliptical disk
model. In the case where the debris streams do not circularize efficiently,
the streams may retain high eccentricities without forming a standard
accretion disk \citep{2017MNRAS.tmp..118S}. In this model, streams may plunge
directly into the black hole when some of the angular momentum is removed,
without losing any energy. Lastly, the outer radius of the elliptical disk
derived by \cite{2018arXiv180802890H} is on the order of $15,\!000\,r_g$,
which is almost two orders of magnitude greater than the self-intersection
radius of a non-spinning black hole with $\Mbh = 10^{6.9}\,\Msun$
\citep[e.g.][]{2017MNRAS.471.1694W}.

Overall, we do not find it necessary to invoke the elliptical disk model to
interprete the spectral line shape. Our spherical outflow model provides a
natural explanation for the flat-topped line shape. Interestingly, the maximum
outflow velocity $v_\text{max} \sim 1.75 \times 10^{4}$~\kms\ employed by our
model is consistent with the velocity of the BAL system. This coincidence may
be explained if the absorption lines are produced by this spherically
expanding material at its outer edge when it scatters photons outside our line
of sight.

\subsection{High Velocity Transient BAL System}

Observations of AT2018zr indicate the presence of the first transient LoBAL
system in a TDE, in which both high- and low-ionization absorption lines are
present. In addition, this transient LoBAL system contains intrinsic hydrogen
Balmer and He\,{\sc i}$^*$ absorption features. It turns out that such systems
are very rare even in QSOs. Blueshifted, broad high-ionization absorption lines
are seen in about 10--20\% of the optically-selected QSOs
\citep{2003AJ....125.1784H,2003AJ....125.1711R,2006ApJS..165....1T,2007ApJ...665..990G}
and are often attributed to accretion disk outflows. However, only about 15\%
of the BALQSOs also show low ionization broad absorption lines (LoBALs) such
as \MgII. A small fraction (15\%) of the LoBAL QSOs that show Fe II or Fe III
absorptions, such as that in Mrk 231, are dubbed as FeLoBALs
\citep[e.g.][]{Veilleux2016}. Since AT2018zr lacks common FeII absorptions in
the NUV, it does not fit in the FeLoBAL category. Currently, there are only a
handful of BALQSOs that exhibit hydrogen Balmer absorption lines
\citep[e.g.][]{2010PASJ...62.1333A, 2012RAA....12..369J, 2015ApJ...815..113Z,
2016ApJ...819...99S, 2017ApJ...838...88S, 2018ApJ...853..167S}

Although Balmer absorption lines are often seen in galaxy spectra, it is
unlikely that the ones we see in AT2018zr are due to intervening galaxies or
clouds because of the following reasons.
First, high-ionization absorption troughs are detected at the same velocities
as the H Balmer absorption lines, suggesting a dynamical association between
them. Second, multi-epoch optical spectra captured the appearance and
disappearance of the H Balmer absorptions on a timescale of days, which is
unexpectedly short if the absorption were due to intervening gas.
Third, the \HeIs\ line must arise from the metastable 2 $^3S$ level, which is
mainly populated by recombination of singly ionized helium ions that requires
a significant amount of photons with $E > 24.6$~eV
\citep{1985ApJ...288..531R}. \HeIs\ is not commonly seen in the ISM since the diffuse
stellar background offers too few photons that can ionize helium and too many
with $E > 4.8$~eV that can ionize the electron in its metastable state.
Lastly, since the TDE is relatively nearby, we should be able to resolve the
intervening galaxies with imaging yet we do not find any.

\subsubsection{Outflows}

Following a star's disruption, the mass initially falls back at a
super-Eddington rate and gradually decreases below the Eddington rate.
Super-Eddington accretion is capable of driving of powerful outflows with
radiation pressure \citep[e.g.][]{2003MNRAS.345..657K,2005ApJ...628..368O}.
\cite{2011MNRAS.415..168S} made the first predictions of the spectroscopic
signatures of super-Eddington outflows in a TDE environment. Their predicted
spectrum is characterized by broad, blueshifted ($v_\text{wind} \sim 0.01$$c$
to 0.1$c$.) absorption features in the super-Eddington phase. Most of these
absorption features are high-ionization lines ($\chi_\text{ion} > 13.6$~eV)
because of the assumption of a hot continuum with a temperature $T_\text{bb}
\gtrsim 10^5$~K. Since the velocity and density of the outflow is viewing
angle-dependent, their model is also able to produce spectra with more NUV and
optical absorption lines, such as \MgII\ and the hydrogen Balmer lines, as a
result of softer continuum. In general, their model predicts a spectrum
similar to that of a BALQSO hence is similar to our observations of AT2018zr.
However, they predict a rapid photospheric temperature evolution and a steep
$t^{-95/36}$ \citep{Strubbe2009,2011MNRAS.410..359L} decline in optical flux
which is not seen in the data.

More recent work has refined the theoretical understanding of how TDEs might
be associated with wide-angle outflows. \cite{2015ApJ...805...83M} suggested that
winds launched by radiation pressure on absorption lines may remove material
as it is drawn in through an extended disk, which would lead to a range of
outflow velocities. \cite{2016MNRAS.461..948M} studied mass-loaded winds launched
from both accretion luminosity and energy released during debris
circularization. They found minimum outflow velocities of $\sim 10,000$ km/s,
similar to the velocities measured here. They also point out that a shell of
promptly launched material will surround the wind consisting of material that
has fallen back more recently. Colder material at the edge of this shell might
provide the origin of the blueshifted absorption troughs seen in the optical
specra of this event.

Another approach has been to perform hydrodynamic simulations of
super-Eddington accretion flows, accounting for the effects of radiation
pressure, magnetic fields, and general relativity, and apply these findings to
TDEs \citep{Sadowski2014,Jiang2017,2018ApJ...859L..20D,Curd2019}. In
particular, \citet{2018ApJ...859L..20D} emphasized the viewing-angle
dependence of the outflow and its resulting emission. While material ejected
in the polar direction had relativistic velocity and was transparent to
X-rays, material launched closer to mid-plane had slower velocity due to
mass-loading, and X-ray emission was highly suppressed along those lines of
sight. The results for such a viewing geometry are consistent with many
aspects of this event, although the time-dependent behavior of this model
requires further study.

The simplest super-Eddington outflow models assume that the wind velocity is a
near-unity fraction of the escape velocity of the gas from the position where
it is launched \citep{Strubbe2009, 2011MNRAS.410..359L}. This would require
the outflowing gas to be launched from a radius of $\approx \,10^{15}$~cm to
match the observed velocity of 0.05$c$ for the absorption features, though the
black hole mass estimation may be somewhat uncertain. However, this radius
derived from the velocity of the absorption features appears to be in tension
with the blackbody radius derived from the UV and optical photometry. The
blackbody radius is found to be steadily decreasing from $10^{15.1}$~cm near
peak to $\lesssim\,10^{14}$~cm at $\Delta t \sim 250$~d
\citep{2019apJ...872..198V}. In this picture, the outflow would be launched at
a radius slightly larger than that traced by the blackbody radius in AT2018zr
when we first observed the \Hb\ absorption on day 30. Meanwhile, our modeling
of the emission lines requires the continuum photosphere to be located within
the edge of the outflow. Accounting for deviations from a pure blackbody
emitter as appropriate for a scattering-dominated atmosphere would only move
the thermalization radius of the continuum to smaller radii
\citep[e.g.][]{Roth2016}, exacerbating the problem. A natural solution is for
mass-loading to reduce the final velocity of the outflow \citep{2016MNRAS.461..948M},
allowing it to be originally launched from smaller radii while matching the
velocity seen in the absorption features.

PS1-11af is the only other TDE in a quiescent galaxy that exhibits transient
broad absorption features (in that case \MgII) with a comparable blueshift
(13,000~\kms) to AT2018zr \citep{Chornock2014}. The velocity of the absorbing
material in PS1-11af is consistent with an outflow launched near the blackbody
radius \citep{Chornock2014}. On the other hand, the line width of \MgII\ (FWHM
$\approx$\,10,200~\kms) in PS1-11af is much broader than the absorption lines
in AT2018zr (FWHMs $\sim$\,3,000~\kms\ for low-ionization lines and
$\sim$\,10,000~\kms\ for high-ionization lines). The broad absorption lines are
expected in the super-Eddington phase due to the wide range of gas velocities
present in the outflow. This gives rise to a linewidth that is similar to the
wind velocity \citep{2011MNRAS.415..168S}. The fact that the absorption lines
in AT2018zr are narrower and detached from their emission peaks may be a
viewing angle effect. For example, a few BALQSOs have been reported with
detached absorption troughs \citep[e.g.\ PG\
1254+046;][]{1998ApJ...500..798H}. This may happen when our line of sight does
not align with the wind such that the materials have already been
substantially accelerated when it intersects our line of sight.


Although many TDE spectroscopic features resemble that of BALQSOs, at least a
fraction of the BALQSOs have low Eddington ratio \citep[$L/L_\text{Edd} <
0.1$;][]{2015AJ....149...85G}, which must have a different mechanism to drive
strong outflows. Indeed, BALQSO wind is thought to be driven by line opacity
in a partially ionized gas. In the line-driven wind framework proposed by
\cite{1995ApJ...451..498M}, dense gas that shields the soft X-ray near the hot
QSO ionizing source is required to avoid overionization and drive the BAL
outflows. Interestingly, the analysis of X-ray observations of AT2018zr also
suggests some X-ray obscuring material residing outside the TDE photosphere
\citep{2019apJ...872..198V}. It is therefore also possible that the LoBAL in
AT2018zr is powered by line opacity, though most BALQSOs are associated with
black holes that are too massive (\Mbh\,$>$\,$10^8$~\Msun) to tidally disrupt
solar-type stars \citep{2011MNRAS.415..168S}.

Despite the TDE sample with UV spectroscopic observations being still small,
the fraction of TDEs that exhibit BAL features in the UV (3 out of 4; \S5.1)
seems to be higher than the fraction ($\sim$\,20\%) of BALQSOs, which is often
attributed to the orientation effect in disk winds. This may suggest that the
TDE outflows have different geometries than QSO outflows and are less
sensitive to viewing angle. Future \hst\ observations of TDEs are desirable for
verifying the large fraction of BALs in TDEs.

\subsubsection{Unbound Debris}

Although the blueshifted absorption features are known to form in disk winds,
we also consider the unbound stellar debris as a possible absorber. In a tidal
disruption, about half of the disrupted star gains enough energy to escape the
black hole on a hyperbolic trajectory, reaching a terminal velocity of
11,000~\kms\ \citep{2016ApJ...827..127K}. Simulations by \cite{Strubbe2009}
showed that the unbound debris irradiated by the accretion disk will produce
emission lines at the optical-IR wavelengths. It is therefore possible that
the rapid evolution in density ($\rho$) and column density ($N$) of the
unbound debris, where $\rho \propto t^{-3}$ and $N \propto t^{-2}$, is causing
the variability of the Balmer absorption seen on the timescale of a few
days.

\begin{figure}[htb]
\centering
\includegraphics[width=3.5in,angle=0,keepaspectratio]{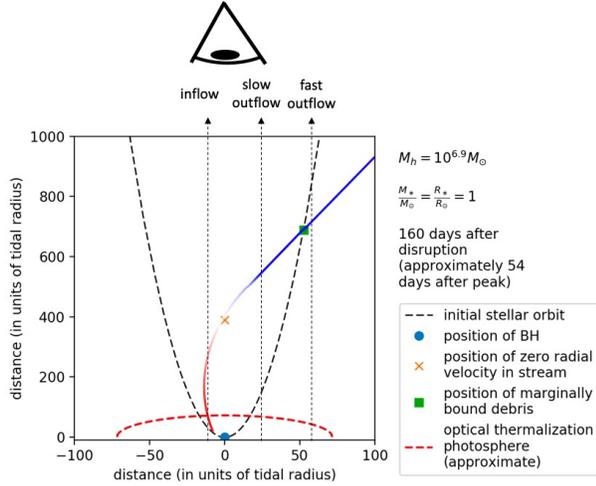}
\caption{A to-scale sketch of the stellar debris at the time corresponding to
the spectral outflow signature.  The red and blue curves trace out the
outflowing and inflowing stellar debris streams, respectively. The viewing
geometry is chosen to allow the unbound debris to provide the high-velocity
absorption lines. However, in such a geometry, a wide range of gas velocities
in the debris stream would intercept the optical photosphere, so we ultimately
disfavor the debris stream as the origin of the high-velocity absorption
lines. In order to identify features of this diagram more clearly, we have
stretched the horizontal distance axis by a factor of 5. We have referred to
\citet{Coughlin2016} to determine the radial location of the portion of the
stream at zero radial velocity.}
\label{fig:los}
\end{figure}

However, the orientation required for an observer to see the high-velocity
unbound debris stream, as depicted in \autoref{fig:los}, will also cover the
slower portions of the stream, as well as the infalling bound stellar debris.
The emergent spectra should therefore have braoder linewidths and even inflow
signatures. However, these are clearly inconsistent with our obervations,
where the photosphere is eclipsed by material at radial velocity of
15,500~\kms\ with a small dispersion of only $\sim$\,3,000~\kms.

\subsubsection{Physical Conditions}

Through out the monitoring period, no significant acceleration or
deacceleration in the absorbing material is detected given the fact that the
absorption features did not show any significant velocity offset between
different epochs. Assuming the absorbing material has been traveling at
constant velocity ($v = 15,\!500$~\kms) since peak light ($t_\text{peak}$), it
would have reached a distance of $\sim$\,$5 \times 10^{15}$~cm by the time we
first observed the \Hb\ absorption on $\Delta t\sim 30$~day. This distance is
just outside of the continuum photosphere estimated from the blackbody fit
\citep{2019apJ...872..198V}.

Observations have revealed a handful of QSOs with hydrogen Balmer and/or
metastable helium lines in their BAL systems. Comparisons of these
observations with photoionization models have constrained the ionization
parameters and densities of the outflows with the photoionization code
\texttt{CLOUDY}, which in turn can constrain the kinetic luminosities of the
outflows.  Below we summarize the physical conditions that give rise to the
hydrogen Balmer and helium lines based on these works, though we remind the
readers to take these values with a grain of salt as the ionizing continuum is
likely to be different for TDEs and AGNs.

Strong helium transitions from the metastable 2s level occur at $\lambda$3889
and $\lambda$10830, which correspond to transition from the metastable state
to 3$p$ and 2$p$ states, respectively. Although once considered a rare
phenomenon, \cite{2015ApJS..217...11L} have found a high association rate
between \HeIs\ and the \MgII\ doublet  among BALQSOs with high S/N spectra
(93\% at S/N\,$\gtrsim$\,35). \cite{2011ApJ...728...94L} have demonstrated
that the \HeIs\ and \HeIss\ lines can serve as a powerful diagnostic of high
column density outflow. This is due to the large $\lambda f_{ik}$ ratio of
23.3, which makes the HeI$^*$ lines sensitive to a wide range of high column
densities ($\tau \propto N_\text{ion}\lambda f_{ik}$) before saturating.
Unfortunately, we are unable to carry out similar analysis due to the \HeIss\
line falling outside our spectroscopic coverage.
%
The presence of \HeI\ provides additional constraints on the ionized gas.
Unlike \HeIs, \HeI\ arises from an energy state that is not metastable, which
is readily depopulated by permitted radiative decays via the $\lambda$10830
transition when the gas density is low. Photoionization calculations have
shown that a gas density of $\log ( n_H / \text{cm}^{-3}) \gtrsim 7$ and an
ionised column density of $\log ( N_H /\text{cm}^{-2}) \gtrsim 23$ are
required to produce $\tau_0$(5876) $\sim$ 0.1 \citep{hamann2019}.

The blueshift velocity of AT2018zr is the highest amongst QSOs with Balmer
BALs \citep{2015ApJ...815..113Z}. Behind the hydrogen recombination front, the
partially ionized regions that give rise to the H Balmer absorption lines
usually have high densities \citep[typically $n_e \sim
10^{6-9}$~cm$^{-3}$;][and references therein]{2015ApJ...815..113Z}. Our
analysis of the H Balmer absorption lines have shown that even the high-order
hydrogen Balmer transitions are saturated. Photoionization modelling with
\texttt{CLOUDY} suggests that a density of $\log (n_H / \text{cm}^{-3})
\gtrsim 6.5$ and a column density of $\log ( N_H /\text{cm}^{-2}) \gtrsim
23.2$ are required to produce a $\tau_0$(\Hgamma)\,$\gtrsim$\,1.3 regardless
of the ionization parameter $U$ \citep{hamann2019}.

With the above estimation of density and column
density, a cloud size of $10^{16} \text{cm}$ can be inferred from $N_H/n_H$
(assuming $\log (n_H / \text{cm}^{-3}) = 7$ and $\log (N_H /\text{cm}^{-2})
\gtrsim 23$. Assuming a spherical cloud, we estimate a total mass of
$4.4\times10^{-3}$\Msun\ and a kinetic energy of $10^{49}$ ergs. The mass and
kinetic energy of the outflow in AT2018zr are about an order of magnitude
greater than that in ASASSN-14li, which is known to host an outflow with
comparable velocity ($12,000 - 36,000$\kms) as derived
from radio observations \citep{Alexander2016}. Note that radio emission, which
is expected as the outflow shocks against the circumnuclear matter (CNM), is not
detected in AT2018zr at a $3\sigma$ upper limit of 10$^{37}$ erg~s$^{-1}$ at 10~GHz
\citep{2019apJ...872..198V}. While the \hst\ data show unambiguous evidence for
the presence of an outflow, the non-detection at radio frequency may be due to the radio signal being buried
by the dense CNM.

Detailed, accurate characterization of the broad emission and absorption lines
in the UV is necessary for robust estimatation of the transient absorber
properties (e.g.\ column density). In particular, the column density and
density are critical parameters to measure the kinetic luminosity, which can
be used to examine different TDE outflow models and assess its importance in
the context feedback in galaxy evolution.

\section{Conclusion}
\label{sec:conclusion}

We report the results of our analysis of multi-epoch UV and optical spectra of
the TDE AT2018zr observed between 2018 March and 2018 December. The wide
wavelength coverage of the \hst\ STIS UV spectra ($\sim$\,1150--3000~\AA)
together with the Keck LRIS spectra ($\sim$\,3500--9000~\AA) allows us to
identify highly blueshifted (15,500~\kms, or 0.05$c$) broad absorption lines
in the UV and optical, including the high- and low-ionization transitions seen
in LoBALQSOs.

In this BAL system, we identified the first hydrogen Balmer and metastable HeI
transitions, which are known to be rare in QSOs, in a TDE. We conclude that
this BAL system is more likely the result of a high velocity outflow launched by TDE
accretion flows, instead of the unbound debris. Given the presence of broad UV
absorption lines are more common in TDEs (3 out of 4) than in QSOs, this may suggest that the outflow launching mechanisms in TDEs are less subjective
to orientation effects than that in QSOs.

Our curve of growth analysis shows that even the high-order hydrogen
Balmer absorption lines are saturated and the absorbing material only covers
the background source partially. By comparing with photoionization models for
AGNs, we conclude that the ionized gas must be characterized by high density
and high column based on the detection of \HeIs\ and \HeI. In order to give
rise to the H Balmer absorption, the partially ionized gas behind the hydrogen
recombination front must also have high column densities. More UV and optical
spectroscopic observations of TDEs will allow detailed photoionization
modeling to assess whether TDE outflows provide significant ``feedback'' in
the context of galaxy evolution.

Using radiative transfer model, we show that the emission profile seen in
expanding TDE outflows \citep{2018ApJ...855...54R} is similar to the
flat-topped \Ha\ line shape in our observations. We find that the spherical
outflow model alone, with a maximum velocity of $1.75 \times 10^4$~\kms\ close
to the velocity of the aforementioned BAL, provides a satisfactory fit to the
observed line profile. While the elliptical disk model have  been invoked to
interprete emission lines in several TDEs, it requires stringent conditions in
order to produce the line shape in AT2018zr. Furthermore, the implied disk
size from the elliptical disk model is two orders of magnitude larger than the
stream self-intersection radius, which is difficult to explain from the
perspective of debris stream dynamics.

We report the appearance of narrow \HeII, \Ha, and \Hb\ emission lines in the
late-time optical spectra of AT2018zr. The line widths of these transitions
imply a virial radius on the order of $10^{17}$~cm. We suggest that the
presence of \HeII\ and N\,{\sc iii}\,$\lambda$4640 emission in later observations may be driven by the
temperature increase in the late-time UV and optical continuum.

TDE spectra are thought to be rather featureless hence the spectroscopic
analysis has been focused on the emission lines in the past. We emphasize
that subtle features, such as the highly blueshifted H Balmer absorption lines
in AT2018zr, may also be present in TDE spectra as a result of outflows. High
S/N spectra with broad wavelength coverage is critical for identifying these
absorption features in the UV and optical. We also recommend monitoring future
TDEs with spectroscopy on a weekly basis, since, as we have shown with
AT2018zr, the spectroscopic features---both emission and absorption---are
variable on such timescale.

\section{Acknowledgements}

T.H.\ thanks Fred Hamann for sharing a copy of his draft prior to publication and Jane Dai for helpful discussions. The UCSC team is supported in
part by NSF grant AST-1518052, the Gordon \& Betty Moore Foundation, the
Heising-Simons Foundation, and from a fellowship from the David and Lucile
Packard Foundation to R.J.F. S.V. acknowledges support from a Raymond and
Beverley Sackler Distinguished Visitor Fellowship and thanks the host
institute, the Institute of Astronomy, where this work was concluded. S.G is
supported in part by NSF grant 1454816. S.V. also acknowledges support by the
Science and Technology Facilities Council (STFC) and by the Kavli Institute
for Cosmology, Cambridge. N. B. acknowledges that this work is part of the
research programme VENI, with project number 016.192.277, which is (partly)
financed by the Netherlands Organisation for Scientific Research (NWO). M. R.
S. is supported by the National Science Foundation Graduate Research
Fellowship Program under Grant No. 1842400. A.G.-Y. is supported by the EU via
ERC grant No. 725161, the ISF, the BSF Transformative program and by a Kimmel
award. The research of Y.Y. is supported through a Benoziyo Prize Postdoctoral
Fellowship. Y.Y. thanks support astronomers at La Palma for assisting the WHT
observation in service mode.

Based on observations made with the NASA/ESA Hubble Space Telescope, obtained
from the Data Archive at the Space Telescope Science Institute, which is
operated by the Association of Universities for Research in Astronomy, Inc.,
under NASA contract NAS 5-26555. The results made use of the Discovery Channel
Telescope at Lowell Observatory. Lowell is a private, non- profit institution
dedicated to astrophysical research and public appreciation of astronomy and
operates the DCT in partnership with Boston University, the University of
Maryland, the University of Toledo, Northern Arizona University, and Yale
University. Some of the data presented herein were obtained at the W.\ M.\
Keck Observatory, which is operated as a scientific partnership among the
California Institute of Technology, the University of California and the
National Aeronautics and Space Administration. The Observatory was made
possible by the generous financial support of the W.\ M.\ Keck Foundation. The
authors wish to recognize and acknowledge the very significant cultural role
and reverence that the summit of Maunakea has always had within the indigenous
Hawaiian community. We are most fortunate to have the opportunity to conduct
observations from this mountain. The William Herschel Telescope is operated on
the island of La Palma by the Isaac Newton Group of Telescopes in the Spanish
Observatorio del Roque de los Muchachos of the Instituto de Astrofísica de
Canarias. The ACAM spectroscopy was obtained as part of OPT/2018A/017. The
results in this work are based on observations obtained with the Samuel Oschin
Telescope 48-inch and the 60-inch Telescope at the Palomar Observatory as part
of the Zwicky Transient Facility project. ZTF is supported by the National
Science Foundation under Grant No. AST-1440341 and a collaboration including
Caltech, IPAC, the Weizmann Institute for Science, the Oskar Klein Center at
Stockholm University, the University of Maryland, the University of
Washington, Deutsches Elektronen-Synchrotron and Humboldt University, Los
Alamos National Laboratories, the TANGO Consortium of Taiwan, the University
of Wisconsin at Milwaukee, and Lawrence Berkeley National Laboratories.
Operations are conducted by COO, IPAC, and UW. SED Machine is based upon work
supported by the National Science Foundation under Grant No. 1106171.

\bibliography{tde}

\end{document}